\documentclass[11pt,a4paper,oneside]{article}
\usepackage[top=3cm, bottom=3cm, left=2cm, right=2cm]{geometry}
\usepackage[english]{babel}
\usepackage[utf8]{inputenc}
\usepackage[T1]{fontenc}

\usepackage{amsthm,amsmath,amssymb,mathrsfs,dsfont,mathtools}
\usepackage{bbm}
\usepackage{bm}
\usepackage{graphicx}
\usepackage[colorlinks=true, allcolors=blue]{hyperref}
\usepackage{comment}
\usepackage[all]{nowidow}
\usepackage{booktabs}
\usepackage{microtype}

\usepackage[authoryear, round]{natbib}
\usepackage{authblk}
\date{}
\title{Bayesian inference for generalized linear models via quasi-posteriors}
\author[1]{Davide Agnoletto}
\author[2]{Tommaso Rigon}
\author[3]{David B. Dunson}
\affil[1]{Department of Statistical Sciences, University of Padova, Via C. Battisti 241, Padova, Italy}
\affil[2]{Department of Economics, Management and Statistics, University of Milano--Bicocca, 20126 Milano, Italy}
\affil[3]{Department of Statistical Science, Duke University, Durham, NC, USA}

\providecommand{\keywords}[1]{
  \small 
  \textbf{\textit{Keywords:}} #1
  \normalsize
}

\usepackage[sf]{titlesec}
\usepackage{sectsty}
\allsectionsfont{\centering \normalfont\scshape} 

\newtheorem{theorem}{Theorem}

\newtheorem{proposition}{Proposition}
\theoremstyle{definition}

\newtheorem{definition}{Definition}

\newtheorem{example}{Example}

\newcommand{\Y}{\bm{Y}}
\newcommand{\y}{\bm{y}}
\newcommand{\X}{\bm{X}}
\newcommand{\x}{\bm{x}}
\newcommand{\bmbeta}{\bm{\beta}}
\newcommand{\hbmbeta}{\Hat{\bm{\beta}}}
\newcommand{\dd}{\mathrm{d}}
\newcommand{\var}{\mathrm{var}}
\newcommand{\E}{\mathds{E}}
\newcommand{\PP}{\mathds{P}}
\newcommand{\R}{\mathds{R}}
\newcommand{\at}[2][]{#1|_{#2}}

\begin{document}

\maketitle

\begin{abstract} Generalized linear models (\textsc{glm}s) are routinely used for modeling relationships between a response variable and a set of covariates. The simple form of a \textsc{glm} comes with easy interpretability, but also leads to concerns about model misspecification impacting inferential conclusions. A popular semi-parametric solution adopted in the frequentist literature is quasi-likelihood, which improves robustness by only requiring correct specification of the first two moments. We develop a robust approach to Bayesian inference in \textsc{glm}s through quasi-posterior distributions. We show that quasi-posteriors provide a coherent generalized Bayes inference method, while also approximating so-called coarsened posteriors. In so doing, we obtain new insights into the choice of coarsening parameter. Asymptotically, the quasi-posterior converges in total variation to a normal distribution and has important connections with the loss-likelihood bootstrap posterior. We demonstrate that it is also well-calibrated in terms of frequentist coverage.  Moreover, the loss-scale parameter has a clear interpretation as a dispersion, and this leads to a consolidated method of moments estimator. 
\end{abstract}


\section{Introduction}

\begin{figure}[t]
\centering
\includegraphics[width=0.85\textwidth]{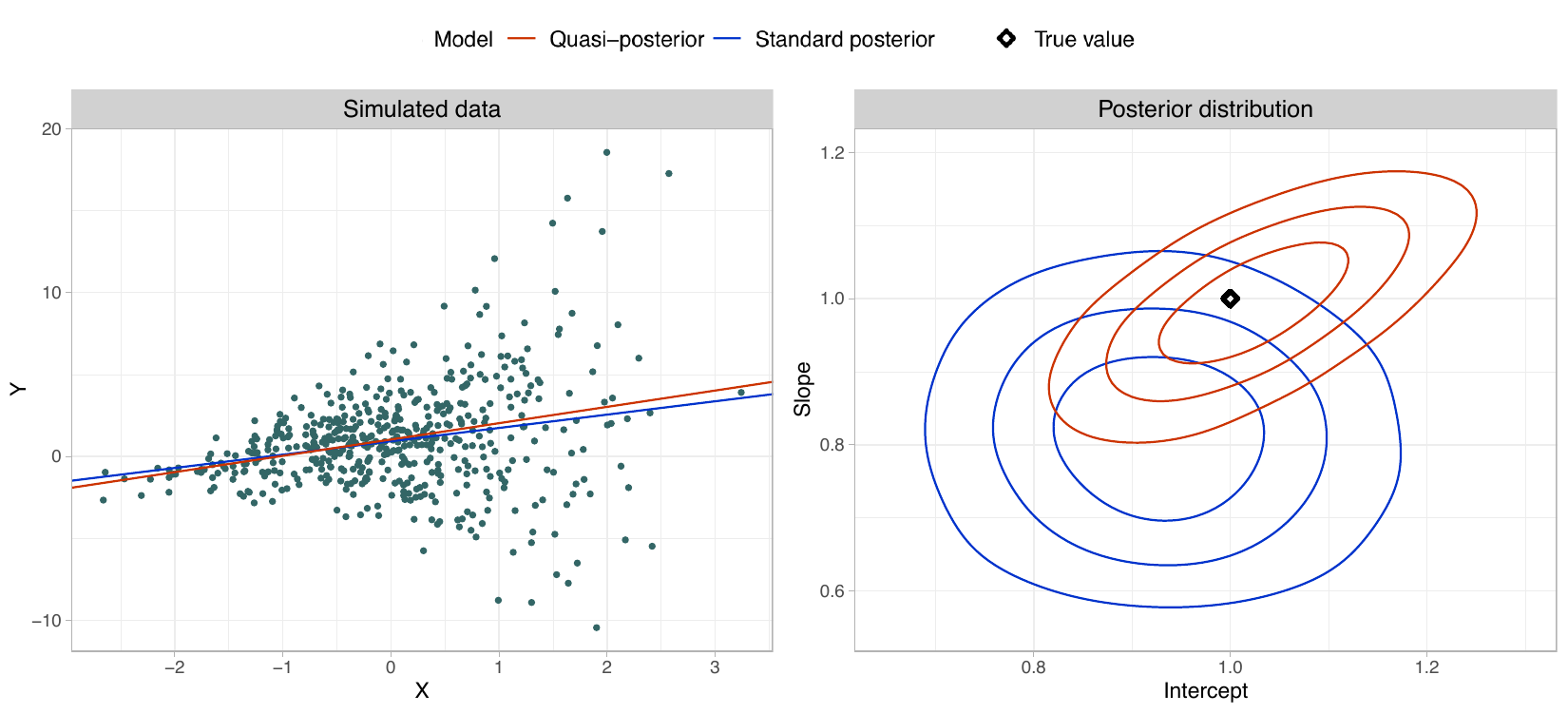}
\caption{\small Comparison of the posterior distributions obtained via Gaussian linear model and quasi-likelihood with $V(\mu_i) = \exp\{\mu_i\}$ on a synthetic dataset of $n = 500$ heteroscedastic observations generated from a Gaussian distribution with variance much larger than the mean.
Left plot: data points.
Right plot: posterior distribution for the two-dimensional regression parameter.}
\label{fig:1}
\end{figure}

Generalized linear models (\textsc{glm}s; \citealp{nelder1972generalized, mccullagh1989generalized}) are a canonical statistical framework for modeling a broad variety of data. Despite their wide use in practice, they often incur  misspecification that could compromise inferential conclusions \citep{muller2013risk, grunwald2017inconsistency}.
For instance, assuming that the distribution of the response variable belongs to a specific exponential family may be untenable in many applications. 
A practical example regards the case of heteroscedastic continuous data, as illustrated in Figure \ref{fig:1}: in this situation the homoscedasticity assumption of the standard Gaussian linear model is wrong, resulting in a posterior which fails to concentrate around the ``true'' parameter value.
Another common case of misspecification occurs when binary or count observations show larger variability than the one assumed by the model. 
This is the so-called overdispersion issue, for which several model-based solutions exist, such as the negative binomial model for count data or the beta-binomial for binary data \citep{gelman2013bayesian}.
However, increased complexity may lead to computational bottlenecks.
Moreover, a model-based solution can result again in misspecification problems.
To bypass these difficulties, one may adopt a nonparametric approach that does not make assumptions about the underlying distribution of the data. On the one hand this allows for greater flexibility, but it can also lead to an increased computational cost and a loss of statistical efficiency and interpretability.

Here we rely on a semi-parametric approach, which represents a midpoint between parametric and nonparametric methods, making weaker assumptions on the distribution of the response variable while preserving  computational tractability. 
The quasi-likelihood method of \citet{wedderburn1974quasi} makes second-order assumptions, meaning that one only needs to specify the mean and variance of the response variable. 
Let $Y_i \in \mathcal{Y} \subseteq \mathds{R}$ denote a response variable and $\bm{x}_i \in \mathcal{X} \subseteq \mathds{R}^p$ be a vector of covariates for $i=1,\ldots,n$. Letting $\bm{\beta} \in \mathds{R}^p$ be the parameter of interest, we assume that
\begin{equation}
\begin{split}
    \mathds{E}(Y_i) & = \mu_i = g^{-1}(\bm{x}_i^\top \bm{\beta}),\\
    \text{var}(Y_i) & = \psi V(\mu_i),
\end{split}
\label{eqn:second_order}
\end{equation}
where $\mu_i\in\mathcal{M} \subseteq \mathds{R}$, $g(\cdot)$ is a strictly monotone and differentiable link function, $V(\cdot)>0$ is a continuous variance function, and $\psi\in(0,\infty)$ is a dispersion parameter. 
The flexibility brought by the second-order assumptions allows the quasi-likelihood to handle the consequences of possible misspecification better than model-based competitors.
Indeed, as shown in the example of Figure \ref{fig:1}, our quasi-posterior adjusts for the substantial bias in the parametric \textsc{glm} posterior, and includes the true parameter value in the center of the distribution.
Within the frequentist literature quasi-likelihood is well-established.
However, in the Bayesian setting such an approach is much more complex to implement for conceptual and practical reasons.
Indeed, within the Bayesian paradigm, a statistician needs to specify all the probabilistic aspects of the model, including those that one may want to leave unspecified. This calls for generalized Bayesian paradigms, which may take into account these practical considerations while maintaining a valid and formal inferential procedure. 

In the last decade, interest in generalized Bayesian methods has grown considerably. 
The paper of \cite{bissiri2016general} represents a cornerstone in this sense, showing that the update of prior beliefs via an exponentiated negative loss in place of the likelihood function is valid and coherent, and hence can be considered a genuine posterior distribution. The proportion of information coming from the loss function and the prior is regulated by a loss-scale parameter.
Its calibration is delicate and, despite several solutions (see \citealp{holmes2017assigning, lyddon2019general}), its choice remains a difficult problem.
Recently, \cite{fong2020marginal} used a generalized Bayes approach to highlight connections between marginal likelihood and cross-validation, while \cite{rigon2023generalized} developed methods for clustering.
\cite{jewson_rossell_2022} provide a general Bayesian selection criterion for the loss function choice based on Hyvärinen score.
For other examples of generalized posteriors using loss functions, see \cite{chernozhukov2003mcmc, royall2003interpreting, zhang2006information, jiang2008gibbs, yin2009gmm}.

Over the years a broad variety of generalized posteriors have been proposed based on different notions of generalized likelihood functions.
These include partial likelihoods \citep{raftery1996, sinha2003, kim2009, ventura2016pseudo}, composite likelihoods \citep{smith2009, pauli2011, ribatet2012}, restricted likelihoods \citep{hoff2007}, and empirical likelihoods \citep{lazar2003}.
These proposals can be placed within the framework of \cite{bissiri2016general} by viewing the log generalized likelihood as a loss.
While the generalized Bayes paradigm focuses on Bayesian updating without the requirement of a true model, the coarsened posterior (c-posterior) framework by \cite{miller2018robust} formalizes the problem of having a small perturbation between the assumed model and the distribution of the observed data. Finally, \cite{grunwald2017inconsistency} propose a tempered likelihood solution for the misspecification problem in linear models.

In this paper, we focus on Bayesian modeling of the first two moments of the unknown data generator without making any specific distributional assumption. 
To do so, we rely on the second-order conditions \eqref{eqn:second_order} and we update the prior beliefs on the parameter of interest through a generalized Bayes approach. More precisely, inference will be based on the quasi-posterior distribution
\begin{equation}\label{eq:gposterior}
    p_\textsc{q}(\bmbeta\mid\y,\X,\psi) \propto p(\bm{\beta})\exp\left\{\sum_{i=1}^n \int_{a}^{\mu_i(\bm{\beta})} \frac{y_i - t}{\psi V(t)} \text{d}t \right\},
\end{equation}
where $\bm{y} = (y_1, \ldots, y_n)^\top$ is the vector of observed responses, $\X = (\x_1^\top, \ldots, \x_n^\top)$ the matrix of covariates, $a\in\mathcal{M}$ is an arbitrary constant that does not depend on $\bmbeta$, and $p(\bm{\beta})$ the prior distribution for the parameter of interest.
There is a literature on related formulations to the quasi-posterior \citep{lin2006quasi, greco2008robust, ventura2010, ventura2016pseudo}. 
However, our work provides novel contributions in formalizing a precise generalized Bayes theory, showing connections with Gibbs posteriors 
\citep{bissiri2016general} and c-posteriors 
\citep{miller2018robust}, 
and proving several new asymptotic properties.
Moreover, although the starting assumptions are similar, the quasi-posterior \eqref{eq:gposterior} is conceptually and practically different from the Bayesian generalized methods of moments of \cite{yin2009gmm}.

Our quasi-posterior takes the form of the c-posterior of \cite{miller2018robust} 
when $V(\mu_i)$ in 
\eqref{eqn:second_order} 
 corresponds to the variance function of an exponential family model. In this case, our quasi-posterior based on second-order assumptions is equivalent to a fully specified model affected by small perturbations. This result provides a new perspective on the coarsening parameter of \cite{miller2018robust}, which we show to be directly connected with the dispersion parameter $\psi$.  This insight has relevant practical implications: c-posteriors have become a popular approach for robustifying Bayesian inferences to model misspecification but there are only heuristic algorithms available for choosing the amount of coarsening. 
In contrast, we propose a simple method of moments estimator.
Even when the above variance function condition is violated, the quasi-posterior is still interpretable as a Gibbs posterior and we can use the same method of moments estimator in choosing the $\psi$ parameter.
Moreover, the quasi-posterior has desirable asymptotic properties, since it converges to a normal distribution centered around the true parameter value and shows correct frequentist coverage asymptotically.

\section{Bayesian inference for GLMs under second-order assumptions}

\subsection{Background on quasi-likelihood}

Let $\Y = (Y_1, \ldots, Y_n)^\top \in \mathcal{Y}^n$ be a vector of random variables whose realization is $\y$.
Let $\x_i \in \mathcal{X}$, $i=1,\ldots,n$, be vectors of linear independent covariates such that $\X$ has full rank for any $n \ge p$.
Standard \textsc{glm}s assume that observations $y_i$ are independent realizations of $Y_i\mid \x_i$, whose distribution belongs to the exponential dispersion family.
Here, we rely on the second-order assumptions \eqref{eqn:second_order}, which specify only the first two moments instead of the entire distribution for $Y_i\mid\x_i$.
We remark on the fact that under the second-order conditions we are assuming the existence of a true value for $(\bmbeta,\psi)$, say $(\bmbeta_0,\psi_0)$, and consequently the existence of a true probability measure ${F_0(\mathrm{d}y\mid\x) = F(\mathrm{d}y\mid\x,\bmbeta_0,\psi_0)}$ on $\mathcal{Y}$ and a joint measure $F_0(\mathrm{d}\bm{y}\mid\bm{X})$ on $\mathcal{Y}^n$. However, in this case, we only make assumptions about the form of mean and variance of the distribution.

A frequentist semiparametric solution for inference under second-order assumptions was introduced by \cite{wedderburn1974quasi}.
Let 
\begin{equation}
    \frac{1}{\psi}\sum_{i=1}^n \frac{y_i - \mu_i}{V(\mu_i)} \frac{\partial\mu_i}{\partial\beta_r},
    \label{eqn:quasi_score}
\end{equation}
be a $p$-dimensional combinant, for $r=1,\ldots,p$.
It coincides with the general form of the score function for an exponential dispersion family and it is an unbiased estimating equation.
Moreover, when the derivative of \eqref{eqn:quasi_score} with respect to $\bmbeta$ is a symmetric matrix, it is possible to specify a function reminiscent of a log-likelihood function (known as log-quasi-likelihood) which has \eqref{eqn:quasi_score} as gradient but does not necessarily correspond to any distribution.
In particular, \cite{wedderburn1974quasi} proposed 
\begin{equation}
    \ell_\textsc{q}(\bmbeta; \y, \X, \psi) = \sum_{i=1}^n \ell_\textsc{q}(\bmbeta; y_i, \x_i, \psi) = \frac{1}{\psi} \sum_{i=1}^n \int_{a}^{\mu_i(\bmbeta)} \frac{y_i - t}{V(t)} \dd t.
    \label{eqn:quasi_likelihood}
\end{equation}
Variance function $V(\cdot)$ must be chosen so that each integral in \eqref{eqn:quasi_likelihood} is finite.
Log-quasi-likelihood is equally suitable for both continuous and discrete data.
The integral can be written in closed form for several choices of variance function, including those corresponding to popular distributions such as 
Gaussian, Poisson, and binomial, as shown in Table \ref{tab:quasi_likelihood}.
A detailed discussion of the properties of quasi-likelihood function is contained in \cite{mccullagh1983quasi}.

\begin{table}[t]
\begin{center}
\begin{tabular}{ccc}
$V(\mu)$ & $\ell_\textsc{q}(\mu;y,\psi)$ & Restrictions\\
\midrule
$1$ & $ \frac{1}{\psi}(y\mu-\frac{1}{2}\mu^2)$ & $-$ \\
$\mu$ & $\frac{1}{\psi}(y\log\mu - \mu)$ & $\mu>0,\;y\ge0$ \\
$\mu^2$ & $\frac{1}{\psi}(-y/\mu - \log\mu)$ & $\mu>0,\;y\ge0$ \\
$\mu^p$ & $\frac{1}{\psi}\big[(1-p)^{-1} y\mu^{1-p} - (2-p)^{-1}\mu^{2-p}\big]$ & $p>2,\;\mu>0,\;y\ge0$ \\
$e^\mu$ & $\frac{1}{\psi}(\mu-y+1)e^{-\mu}$ & $-$ \\
$\mu(1-\mu)$ & $\frac{1}{\psi}\big[ y\log\mu + (1-y)\log(1-\mu)\big]$ & $\mu \in (0,1),\;y\in[0,1]$ \\
$\mu^2(1-\mu)^2$ & $\frac{1}{\psi}\big[(2y-1)\log\frac{\mu}{1-\mu} + \frac{2y-1}{1-\mu} - \frac{y}{\mu(1-\mu)}\big]$ & $\mu \in (0,1),\;y\in[0,1]$\\
$\mu + \mu^2/k$ & $\frac{1}{\psi}\big(y\log\frac{\mu}{k+\mu} + k\log\frac{1}{k+\mu}\big)$ & $\mu>0,\;y\ge0$\\
\end{tabular}
\caption{Closed form of the log-quasi-likelihood function for several choices of variance functions.}
\label{tab:quasi_likelihood}
\end{center}
\end{table}

\subsection{Quasi-posterior}

In classical Bayesian statistics, inference about $\bmbeta$ is carried out using the posterior distribution, which 
updates the prior distribution with the likelihood function.
Under only the second-order assumptions, without additional conditions, a genuine likelihood is not available. However, a posterior distribution for $\bmbeta$ can be obtained using the quasi-likelihood.
\begin{definition}
Let $\exp\{\ell_\textsc{q}(\bmbeta;\y,\X,\psi)\}$ be the quasi-likelihood function and $p(\bmbeta)$ be the prior distribution for $\bmbeta$.
We refer to
\begin{equation*}
    p_\textsc{q}(\bmbeta\mid\y,\X,\psi) \propto p(\bmbeta) \exp \left\{ \ell_\textsc{q}(\bmbeta;\y,\X,\psi) \right\} = 
    p(\bmbeta) \exp\left\{ \frac{1}{\psi} \sum_{i=1}^n \int_{a}^{\mu_i(\bmbeta)} \frac{y_i - t}{V(t)} \text{d}t \right\}
    \label{eqn:quasi_posterior}
\end{equation*}
\label{def:quasi_posterior}
as the quasi-posterior distribution for $\bmbeta$.
\end{definition}
The quantity $p_\textsc{q}(\bmbeta\mid\y,\X,\psi)$ is a well-defined probability density function provided that the normalizing constant is such that $0 < \int p(\bmbeta) \exp \left\{ \ell_\textsc{q}(\bmbeta;\y,\X,\psi) \right\}\dd\bmbeta < \infty$.
Although this condition is not always true in general, it can be easily verified that it holds for a wide range of variance functions, including those most commonly used in practice.

The quasi-posterior is a rational update of a belief distribution under a generalized Bayesian model 
\citep{bissiri2016general}.
In a regression context, 
the target of inference for a generalized Bayes procedure is $\bmbeta_{\textsc{opt}}$, defined as the minimizer of an integrated loss function,
\begin{equation}
    \bmbeta_{\textsc{opt}} = \arg \min_{\bmbeta} \int \ell(\bmbeta;\bm{y},\X) F_0(\mathrm{d}\bm{y}\mid\bm{X}),
    \label{eqn:beta_optimal}
\end{equation}
where $\ell(\bmbeta;\y,\X)$ is a loss function and $\bmbeta_{\textsc{opt}}$ is assumed to be unique. \cite{bissiri2016general}
used a decision-theoretic approach to determine a coherent update of the prior beliefs about 
$\bmbeta_{\textsc{opt}}$ without knowledge of the likelihood function.
Letting $\mathcal{P}$ be the space of all conditional distributions of $\bmbeta$ given the data, the following loss function is used to measure how well 
$\nu \in \mathcal{P}$ represents posterior beliefs:
\begin{equation*}
    \mathscr{L}\{\nu(\bmbeta)\} = \frac{1}{\psi}\mathds{E}\left\{\ell(\bmbeta;\y,\X)\right\} + \textsc{kl}\{\nu(\bmbeta) \mid \mid p(\bmbeta)\},
\end{equation*}
where the expectation is taken with respect to $\nu(\bmbeta)$ and \textsc{kl} represents the Kullback-Leibler divergence.
The first term quantifies the discrepancy between $\nu$ and the information about the parameter coming from the observed data, while the second quantifies the closeness to the prior.
The loss-scale parameter $\psi>0$ regulates the balance between these two terms.
The only coherent update of the prior beliefs $p(\bmbeta)$ about $\bmbeta_{\textsc{opt}}$, after observing the data, to the posterior $p_\psi(\bmbeta\mid\y,\X)$ is
\begin{equation}
    p_\psi(\bmbeta\mid\y,\X) = \arg\min_{\nu} \mathscr{L}\{\nu(\bmbeta)\} \propto p(\bmbeta)\exp\left\{-\frac{1}{\psi} \ell(\bmbeta;\y,\X)\right\},
    \label{eqn:bissiri_optimal}
\end{equation}
where $\ell(\bmbeta;\y,\X) = \sum_{i=1}^n\ell(\bmbeta;y_i,\x_i)$ is an additive loss function.
It is straightforward to see that the quasi-posterior of Definition \ref{def:quasi_posterior} corresponds to the minimizer \eqref{eqn:bissiri_optimal} with 
\begin{equation}
    \ell(\bmbeta;\y,\X) = - \sum_{i=1}^n \int_{a}^{\mu_i(\bmbeta)} \frac{y_i-t}{V(t)}\dd t.
    \label{eqn:loss_function}
\end{equation}

Taking a generalized Bayesian approach, the target of quasi-posterior inference is $\bmbeta_{\textsc{opt}}$ defined in~\eqref{eqn:beta_optimal}.
The second-order conditions assume the existence of a true value $\bmbeta_0$ for the regression parameter.
The relation between this two quantities is clarified by the following Theorem.
\begin{theorem}
\label{th:beta_opt_zero}
    For quasi-posteriors, suppose the target of inference $\bmbeta_{\textsc{opt}}$ defined in equation~\eqref{eqn:beta_optimal} is unique. Then, it must coincide with the true value $\bmbeta_0$ assumed under the second-order conditions
    \begin{equation*}
        \bmbeta_{\textsc{opt}} = \arg\max_{\bmbeta}\int\ell_Q(\bmbeta;\y,\X,\psi)F_0(\mathrm{d}\y\mid\X) = \bmbeta_0.
    \end{equation*}
\end{theorem}
The crucial assumption of the \cite{bissiri2016general} framework is that $\bmbeta_{\textsc{opt}}$ should represent a quantity of practical interest.
Theorem \ref{th:beta_opt_zero} 
shows that this is indeed the case for our quasi-posterior - the quantity targeted in the generalized Bayesian paradigm coincides with the true parameter value. Hence, the quasi-posterior represents a worthwhile representation of 
posterior beliefs about $\bmbeta$.

\section{Connection with c-posteriors}
\label{sec:c-posteriors}

C-posteriors provide an alternative approach for robustifying Bayesian inferences to model misspecification. In this section, we formally connect our quasi-posterior with c-posteriors, which are briefly reviewed here. Start by 
choosing a parametric model for the data generator such that $Y_i\mid\x_i$, $i=1,\ldots,n$, are drawn from the density function $f(y \mid \bmbeta,\x)$.
The observed responses $\y$ are assumed to be perturbed realizations of $\Y$, in the sense that ${d(\Y,\y\mid\X)<R}$ for some measure of discrepancy $d(\cdot,\cdot\mid\cdot)$ and some $R \in (0,\infty)$. 
Suppose that the observed data $y_i\mid\x_i$, $i=1,\ldots,n$, behave like independent samples from some unknown perturbed density $\Tilde{f}(y\mid\x)$.
Inference on $\bmbeta$ using the standard posterior is equivalent to conditioning on the event $\Y=\y$. However, this is incorrect when there is a perturbation.
With this motivation, \cite{miller2018robust} propose to infer $\bmbeta$ conditioning on the event $d(\Y,\y\mid\X)<R$.
The resulting posterior distribution is the so-called c-posterior, defined as ${p\left( \bm{\beta} \mid d(\bm{Y}, \bm{y} \mid \bm{X}) < R\right)}$.
It is important to clarify we are not assuming that $\Tilde{f}(\dd y\mid\x)$ corresponds to the density $f_0(y\mid \x)$ of $F_0(\mathrm{d} y\mid \x)$: the first is the perturbed law that generates the observations according to \cite{miller2018robust} approach, while the second is the unknown true mechanism of which we are assuming the first two moments.

A reasonable choice of $d(\cdot,\cdot\mid\cdot)$ is the estimator of the conditional relative entropy between $\Tilde{f}( y\mid\x)$ and $f(y \mid \bmbeta,\x)$, that is
$d(\Y,\y\mid\X) = n^{-1} \sum_{i=1}^n \big[\log\Tilde{f}(y_i\mid\x_i)-\log f( y_i\mid\bmbeta,\x_i)\big]$.
Then, if $R \sim \text{Exp}(\alpha)$, the regression c-posterior can be approximated with a power posterior of the form
\begin{equation*}
   p_{\alpha}(\bm{\beta} \mid \bm{y}, \bm{X}) \propto p(\bm{\beta}) \prod_{i=1}^n f(y_i \mid\bmbeta,\x_i)^{\frac{\alpha}{\alpha+n}},
   \label{eqn:power_posterior}
\end{equation*}
where $\alpha/(\alpha+n)$ represents the amount of coarsening.
The choice of $\alpha$ is crucial since it regulates the amount of coarsening and consequently the concentration of the posterior: intuitively, to raise a likelihood to the power of $\alpha/(\alpha+n)$ corresponds to adjusting the sample size from $n$ to $n\alpha/(\alpha+n)$, so the c-posterior is as concentrated as a regular posterior having sample size  $n\alpha/(\alpha+n)$.
When $\alpha \gg n$, the power posterior will be similar to the standard posterior, 
while the amount of coarsening tends to $\alpha/n$ if $\alpha\ll n$.
\cite{miller2018robust} propose heuristics for choosing $\alpha$, including simply letting $\alpha = n$, and a graphical criterion based on a calibration curve which compares, for a fixed $\alpha$, the posterior expected log-likelihood, chosen as a measure of goodness of fit, and the posterior expected model complexity. 
However, current approaches for choosing $\alpha$ lack theoretical support.

A solution can be found in the connection between quasi- and c-posteriors.
Indeed, the power posterior approximation of the c-posterior is recovered as a particular case of the quasi-posterior when the assumed model is an exponential family.
The crucial point is the relation between the quasi-likelihood and the genuine likelihood for the exponential family, which is recalled in the following Theorem.
\begin{theorem}[\citealp{wedderburn1974quasi}]
\label{th:wedderburn}
For a single observation  $Y$, the log-likelihood function has score function $(y-\mu)V(\mu)^{-1}$, where $\mu = \mathds{E}(Y)$ and $V(\mu) = \var(Y)$, if and only if the density of $Y$ with respect to some measure can be written in the exponential form $\exp\{y\theta - b(\theta)\}$, where $\theta = \theta(\mu)$ is the so-called natural parametrization.
\end{theorem}
In other words, the genuine log-likelihood of an exponential family coincides with the corresponding log-quasi-likelihood, up to a multiplicative factor.
Table \ref{tab:quasi_likelihood} contains the examples for Gaussian, Poisson, and binomial distributions corresponding to $V(\mu)=1$, $V(\mu)=\mu$, and $V(\mu)=\mu(1-\mu)$, respectively.
The connection between quasi- and c-posterior arises naturally and is summarized in the following Proposition.
\begin{proposition}
\label{prop:cposterior}
Assume the conditional distribution of $Y_i \mid \x_i$, $i=1,\ldots,n$, to be an exponential family of order 1, that is
\begin{equation}
    f(y_i \mid \bmbeta,\x_i) = h(y_i)\exp\left\{\theta(\mu_i)y_i - b(\theta(\mu_i))\right\},
    \label{eqn:exp_fam}
\end{equation}
where $\theta(\cdot)$ is the natural parametrization, $\mu_i = \mu_i(\bmbeta) = g^{-1}(\x_i^\top\bmbeta)$, and $h(y_i)$ the normalizing constant.
If $V(\cdot)$ in \eqref{eqn:second_order} corresponds to the variance function of \eqref{eqn:exp_fam}, then the corresponding quasi-posterior is equivalent to the power posterior approximation of the c-posterior based on \eqref{eqn:exp_fam}, with $\psi = (\alpha+n)/\alpha$.
\end{proposition}
It follows directly that the quasi-posterior is approximately proportional to the regression c-posterior distribution based on conditional relative entropy.
This relation links posteriors based on two different notions of model misspecification:
using log-quasi-likelihood as a loss function within a Gibbs posterior \citep{bissiri2016general} only requires specification of the first two moments, while the c-posterior requires a complete specification of the likelihood of the data but allows this likelihood to be slightly misspecified. The above Proposition unifies these two approaches for robust Bayesian inference for \textsc{glm}s.
This unification highlights the relation between amount of coarsening and dispersion in the data, allowing estimation of $\alpha$ based on $\psi$ as discussed in Section \ref{sec:connection_llb}.
The following Example clarifies the link between $\alpha$ and $\psi$ for count data.

\begin{example}[Overdispersed Poisson regression]
\label{ex:quasi_poisson}
    Consider the relative entropy regression c-posterior for count response with the assumed model a Poisson distribution with canonical link function. The corresponding power posterior approximation is
    \begin{equation*}
        p_{\alpha}(\bm{\beta} \mid \bm{y}, \bm{X}) \propto p(\bm{\beta}) \exp\left\{\frac{\alpha}{\alpha+n}\sum_{i=1}^n \left(y_i\x_i^\top\bmbeta - \exp\{\x_i^\top\bmbeta\}\right)\right\}.
    \end{equation*}
    Assuming $\E(Y_i) = \exp\{\x_i^\top\bmbeta_0\}$ and $\var(Y_i) = \psi\exp\{\x_i^\top\bmbeta_0\}$, the quasi-likelihood \eqref{eqn:quasi_likelihood} is $\ell_\textsc{q}(\bmbeta;\y,\X,\psi) \propto \psi^{-1}\sum_{i=1}^n\big(y_i\x_i^\top\bmbeta - \exp\{\x_i^\top\bmbeta\}\big)$, which leads to the quasi-posterior
    \begin{equation}
        p_\textsc{q}(\bmbeta\mid\y,\X,\psi) \propto p(\bm{\beta}) \exp\left\{\frac{1}{\psi}\sum_{i=1}^n \left(y_i\x_i^\top\bmbeta - \exp\{\x_i^\top\bmbeta\}\right)\right\}.
        \label{eqn:quasi_poisson_posterior}
    \end{equation}
    The two posteriors are equivalent when 
    \begin{equation*}
        \alpha = \frac{n}{\psi-1}.
    \end{equation*}
This relation offers some insights into the phenomenon that c-posteriors and quasi-posteriors can be robust against.
In this example, $\psi$ controls the difference between the dispersion of the observations and the variance assumed by the parametric model: for $\psi = 1$ the Poisson model is well specified, while if $\psi>1$ or $\psi\in(0,1)$ it is affected by overdispersion or underdispersion, respectively.
Both the quasi-posterior and c-posterior are robust against overdispersion, since $\psi>1$ implies $\alpha\in(0,\infty)$, and they recover the standard posterior when $\psi\rightarrow1$ and so $\alpha\rightarrow\infty$.
However, the quasi-posterior is also robust against underdispersion while the c-posterior is not, since $\alpha>0$ by construction.
\end{example}

\section{Asymptotic properties}
\label{sec:asymptotic_properties}

In this section we study large sample properties of quasi-posteriors. The literature contains some related work on  generalized posteriors; for example, studying convergence in distribution (see \citealp{greco2008robust, ventura2016pseudo}) or in total variation (\citealp{chernozhukov2003mcmc}) to a normal distribution.
Here we provide results on limiting distributions and 
frequentist coverage of quasi posteriors.
In the following theorem, based on Theorems 4 and 5 of \cite{miller2021asymptotic}, we present the asymptotic behavior of the quasi-posterior distribution.
\begin{theorem}
\label{th:asymptotic}
Let $p(\bmbeta)$ be continuous at $\bmbeta_0$ and such that $p(\bmbeta_0)>0$.
Let $Y_i\mid\x_i$, $i=1,\ldots,n$ be independent random variables  generated according to $F_0(\mathrm{d}y \mid \bm{x})$ and assume second-order conditions of equation~\eqref{eqn:second_order} are satisfied. Let $\hbmbeta=\arg\max_{\bmbeta}\ell_\textsc{q}(\bmbeta;\y,\X,\psi)$ be the empirical risk minimizer and $i(\bmbeta) = \lim_{n\rightarrow\infty}(\psi n)^{-1} \X^\top \bm{D} \X$ where $\bm{D}=\mathrm{diag}(d_1,\ldots,d_n)$, $d_i = [V(\mu_i)g'(\mu_i)^2]^{-1}$, for $i = 1,\ldots,n$. Then, under regularity conditions, the density $p_n(\bmbeta)$ of the centered and scaled quasi-posterior $\sqrt{n}(\bmbeta - \hbmbeta)$ is such that
\begin{equation*}
     \left|p_n(\bmbeta) - \mathcal{N}_p\left(\bmbeta \mid 0, i(\bmbeta_0)^{-1}\right)\right|_\text{TV} \overset{F_0}{\longrightarrow} 0,  \qquad n\rightarrow \infty,
\end{equation*}
almost surely. Moreover, the estimator $\hat{\bmbeta}$ is consistent, that is $\hat{\bmbeta} \rightarrow \bmbeta_0$ almost surely.
\end{theorem}
The regularity conditions are discussed in Appendix~\ref{proof_asymptotics}. Theorem \ref{th:asymptotic} shows that the quasi-posterior converges in total variation to a Gaussian distribution centered around the maximum quasi-likelihood estimator $\hbmbeta$, which is a consistent estimator of $\bmbeta_0$. 
Hence, the quasi-posterior for $\bmbeta$ asymptotically concentrates around the true parameter value.
The quantity $\hbmbeta$ is the solution  of the unbiased estimating equation obtained with \eqref{eqn:quasi_score}, and is an M-estimator.
Furthermore, when \eqref{eqn:quasi_score} coincides with the score function of an exponential family, it is equivalent to the maximum likelihood estimator (\textsc{mle}).
As a consequence of Theorem \ref{th:beta_opt_zero}, $\hbmbeta$ is also the empirical counterpart of $\bmbeta_{\textsc{opt}}$, often called the empirical risk minimizer.
The covariance of the asymptotic distribution is determined by the expected information matrix
\begin{equation}
    I(\bmbeta) = \E\left[-\nabla^2\ell_\textsc{q}(\bmbeta;\Y,\X,\psi)\right]
    = \psi^{-1}\X^\top \bm{D} \X,
    \label{eqn:information_ql}
\end{equation}
where 
$\nabla$ denotes the gradient operator with respect to $\bmbeta$,
computed for $\bmbeta = \bmbeta_0$.
Consequently, $\psi$ governs the concentration of quasi-posterior around $\hbmbeta$: as $\psi$ grows, the asymptotic variance is inflated, forcing the quasi-posterior to be flatter.
Theorem~\ref{th:asymptotic} is stated for a general form of quasi-likelihood, but more specific conclusions for several popular choices of link function and variance function could be obtained, e.g.
\begin{enumerate}
    \item $\E(Y_i) = \exp\{\x_i^\top\bmbeta_0\}$, $\var(Y_i) = \psi_0 \exp\{\x_i^\top\bmbeta_0\}$ (overdispersed Poisson regression);
    \item $\E(Y_i) = \exp\{\x_i^\top\bmbeta_0\}(1+\exp\{\x_i^\top\bmbeta_0\})^{-1} = \mu_i$, $\var(Y_i) = \psi_0\mu_i(1-\mu_i)$ (overdispersed logistic regression);
    \item $\E(Y_i) = \x_i^\top\bmbeta_0$, $\var(Y_i) = \psi_0 \exp\{\x_i^\top\bmbeta_0\}$ (heteroscedastic continuous data regression).
\end{enumerate}

While for the overdispersed Poisson regression and overdispersed logistic regression the result of Theorem~\ref{th:asymptotic} may not be surprising since the derivatives are equivalent to the parametric case up to a constant, it remarkably holds, under mild regularity conditions, even for the case of heteroscedastic continuous data regression.
\begin{example}[Heteroscedastic continuous data regression]
\label{ex:het_qp}
    Assume $\E(Y_i) = \x_i^\top\bmbeta_0$ and $\var(Y_i) = \psi_0 \exp\{\x_i^\top\bmbeta_0\}$, $i=1,\ldots,n$.
    The corresponding quasi-posterior distribution is
    \begin{equation*}
        p_\textsc{q}(\bmbeta\mid\y,\X,\psi) \propto p(\bmbeta) \exp\left\{\frac{1}{\psi}\sum_{i=1}^n (\x_i^\top\bmbeta - y_i +1)\exp\{-\x_i^\top\bmbeta\}\right\}.
    \end{equation*}
    Thus, the quasi posterior distribution of $\bmbeta$ is approximately a $p$-variate Gaussian with mean $\hat{\bmbeta}$ and covariance matrix 
    $I(\bmbeta_0)^{-1}$, with
    $I(\bmbeta_0) = \psi^{-1}\X^\top\Tilde{\bm{D}}\X$ and $\Tilde{\bm{D}}=\mathrm{diag}(\Tilde{d}_1,\ldots,\Tilde{d}_n)$, where $\Tilde{d}_i = \exp\{-\x_i^\top\bmbeta_0\}$, for $i=1,\ldots,n$. The matrix $I(\bmbeta_0)$ can be practically evaluated using the plug-in estimate~$I(\hbmbeta)$.
\end{example}

\subsection{Correct frequentist coverage}
\label{sec:coverage}

A notable property of the asymptotic distribution of the quasi-posterior is correct frequentist coverage.
We say that the quasi-posterior is well-calibrated in terms of frequentist coverage when, asymptotically, the posterior credible sets of level $\rho$ contain the true parameter value $\bmbeta_0$ with probability $\rho$.
Generalized posteriors do not automatically have this property, so quasi-posteriors are special in this regard.
\cite{miller2021asymptotic} provides sufficient conditions for which a generalized posterior is asymptotically calibrated in terms of coverage. Roughly speaking, a generalized posterior for $\bmbeta$ has correct coverage if converges in total variation distance to the same normal distribution 
to which the empirical risk minimizer $\hbmbeta$ converges in distribution.
Although this condition is not satisfied for many common choices of generalized posteriors, it holds for quasi-posteriors.
This allows the quasi-posteriors to have asymptotically correct frequentist coverage, thanks to the general results of \cite{miller2021asymptotic}. The proof is discussed in Appendix~\ref{proof_coverage}, together with a description of the regularity conditions.
\begin{theorem}
\label{th:coverage}
    Let $S_1,S_2,\ldots\subseteq\R^p$ be a sequence of convex credible sets for the quasi-posterior of $\bmbeta$ of asymptotic level $\rho\in(0,1)$. 
    Then, under mild regularity conditions and assuming $\psi = \psi_0$ 
    \begin{equation*}
        \PP(\bmbeta_0\in S_n\mid\y,\X,\psi_0) \rightarrow\rho
    \end{equation*}
    as $n\rightarrow\infty.$
\end{theorem}

\subsection{Connection with loss-likelihood bootstrap}
\label{sec:connection_llb}

The quasi-posterior has connections with the loss-likelihood bootstrap introduced by \cite{lyddon2019general} and their proposal for setting the value of the loss-scale parameter.
The loss-likelihood bootstrap is a version of the Bayesian bootstrap \citep{rubin1981bayesian} which, essentially, enables generation of samples from an approximate posterior distribution built from a loss function.
Asymptotically, the loss-likelihood bootstrap posterior converges to a normal distribution with covariance matrix that does not depend on $\psi$.
Conversely, the asymptotic normal distribution of a generalized posterior having the same loss function depends on the loss-scale parameter; an example is the result of Theorem \ref{th:asymptotic} for the quasi-posterior.
For this reason, \cite{lyddon2019general} proposed to choose $\psi$ by matching the amount of information of the two asymptotic distributions quantified in terms of the Fisher information number, that, for a normal distribution, is asymptotically equal to the trace of the Fisher information matrix.
The matching value is
\begin{equation}
    \psi_{\textsc{llb}} = \frac{\text{tr}\{j(\bmbeta_0)\}}{\text{tr}\{j(\bmbeta_0)h(\bmbeta_0)^{-1}j(\bmbeta_0)\}},
    \label{eqn:psi_lyddon}
\end{equation}
where 
$$j(\bmbeta) = \lim_{n\rightarrow\infty}n^{-1}\E\left[\nabla^2\ell(\bmbeta;\Y,\X)\right],\qquad
h(\bmbeta) = \lim_{n\rightarrow\infty}n^{-1}\E\left[\nabla\ell(\bmbeta;\Y,\X)\nabla\ell(\bmbeta;\Y,\X)^\top\right].$$
This result refers to a generic loss function $\ell(\cdot)$.
However, for quasi-posteriors we obtain a substantial simplification.
\begin{proposition}
    If $\E(Y_i) = g^{-1}(\x_i^\top\bmbeta_0) = \mu_i(\bmbeta_0)$ and $\mathrm{var}(Y_i) = \psi_0 V(\mu_i(\bmbeta_0))$, with both operators taken with respect to $F_0(\dd y_i\mid\x_i)$, and loss function in equation~\eqref{eqn:loss_function}, then $\psi_{\textsc{llb}} = \psi_0$.
    \label{prop:psi_lyddon}
\end{proposition}
This means that under the second order conditions, the value of $\psi$ that equates the Fisher information number of the two asymptotic distributions is the true value $\psi_0$.
In practice, of course, $\psi_0$ is unknown and the common usage in the generalized Bayes literature is to assign a value to $\psi$ based on the data before approximate sampling from the generalized posterior (see \citealp{bissiri2016general, holmes2017assigning,rigon2023generalized}).
For example, \cite{lyddon2019general} propose the estimator ${\Hat{\psi}_{\textsc{llb}} = \text{tr}\{j_n(\hbmbeta)\}\text{tr}\{j_n(\hbmbeta)h_n(\hbmbeta)^{-1}j_n(\hbmbeta)\}^{-1}}$, where ${j_n(\bmbeta) = n^{-1}\sum_{i=1}^n\nabla^2\ell(\bmbeta;y_i,\x_i)}$ and $h_n(\bmbeta) = n^{-1}\sum_{i=1}^n\nabla\ell(\bmbeta;y_i,\x_i)\nabla\ell(\bmbeta;y_i,\x_i)^\top$ are the empirical versions of $j(\bmbeta)$ and $h(\bmbeta)$, respectively.
However, it is not guaranteed that $\Hat{\psi}_{\textsc{llb}}$ is a consistent estimator of $\psi_0$.
For this reason we suggest to use the estimator based on the method of moments
\begin{equation}
    \Hat{\psi} = \frac{1}{n-p}\sum_{i=1}^n\frac{\big(y_i-\Hat{\mu}_i\big)^2}{V\big(\Hat{\mu}_i\big)},
    \label{eqn:mom_psi}
\end{equation}
where $\Hat{\mu}_i = \mu_i(\hbmbeta)$.
Equation \eqref{eqn:mom_psi} produces a consistent estimator of the true dispersion parameter and it represents the standard choice in the frequentist quasi-likelihood literature \citep{wedderburn1974quasi, mccullagh1983quasi, mccullagh1989generalized}.
An exemplification for the comparison of $\Hat{\psi}$ and $\Hat{\psi}_{\textsc{llb}}$ is the Gaussian loss with $\beta\in\mathds{R}$ illustrated in Example \ref{ex:psi}.
\begin{example}
    Let $\E(Y_i) = x_i\beta$ and $\var(Y_i) = \psi$, for $i=1,\ldots,n$.
    Then the method of moments estimator for $\psi$ is the well known variance estimator ${\Hat{\psi} = (n-1)^{-1}\sum_{i=1}^n(y_i-x_i\Hat{\beta})^2}$, while the one based on loss-likelihood bootstrap is ${\Hat{\psi}_{\textsc{llb}} = (\sum_{i=1}^nx_i^2)^{-1} \sum_{i=1}^n x_i^2(y_i-x_i\Hat{\beta})^2}$.
    However, while $\Hat{\psi}$ is unbiased, $\Hat{\psi}_{\textsc{llb}}$ is not, since $\E(\Hat{\psi}_{\textsc{llb}})=\psi\big[1-\sum_{i=1}^nx_i^4 / (\sum_{i=1}^nx_i^2)^2\big]$.
    \label{ex:psi}
\end{example}
A further justification for \eqref{eqn:mom_psi} is provided by the coverage result of Theorem \ref{th:coverage}.
Indeed, a condition for having correct coverage is that the quasi-posterior and empirical risk minimizer have the same asymptotic covariance matrix, which depends on $\psi$.
Since the standard frequentist choice for its estimation is \eqref{eqn:mom_psi}, the method of moments estimator represents the solution that keeps this equivalence valid in practice. 

Importantly, $\hbmbeta$ does not depend on $\psi$. Thus, by estimating $\psi$ using the empirical risk minimizer, we avoid need for an iterative algorithm.
The estimated value can be interpreted as the observed amount of dispersion, based on the value of $\bmbeta$ that minimizes the quasi-likelihood function in the sample.

\section{Illustrations}

Quasi-posteriors are a useful tool in many practical situations.
In this section we illustrate some simulated and real data examples of quasi-posteriors in different contexts where traditional model-based approaches struggle.

\subsection{Simulation example: heteroscedastic continuous data}

To show consequences of model misspecification on estimates of regression coefficients, we apply a Gaussian linear model to heteroscedastic continuous data and compare performance to a quasi-posterior with mean and variance correctly specified.
We produce 100 simulated datasets, each of size $n=300$.
In each simulation, the true data generator for $y_i\mid\x_i$ is $\mathcal{N}(\x_i^\top\bmbeta_0,\psi_0 \exp\{\x_i^\top\bmbeta_0\})$, $i=1,\ldots,n$, with true parameter value $\bmbeta_0=(-3,2,1.5,1)$ and $\psi_0=2.5$.
The covariates $\x_i$ are sampled from independent standard normal distributions.

\begin{table}[t]
\centering
\caption{\small Heteroscedastic continuous data example. Frequentist coverage of credible intervals with nominal level $\rho \in \{0.90, 0.95, 0.99\}$ for Gaussian linear model (\textsc{lm}) and quasi-posterior (\textsc{qp}).}
\label{tab:het_sim_covergage}
\vspace{0.2cm}
\footnotesize
\begin{tabular}{l c cccc c cccc c cccc}
 & & \multicolumn{4}{c}{$\rho = 0.90$} & & \multicolumn{4}{c}{$\rho = 0.95$} & & \multicolumn{4}{c}{$\rho = 0.99$}\\
 & & $\beta_1$ & $\beta_2$ & $\beta_3$ & $\beta_4$ & & $\beta_1$ & $\beta_2$ & $\beta_3$ & $\beta_4$ & & $\beta_1$ & $\beta_2$ & $\beta_3$ & $\beta_4$ \\
\textsc{lm} & & 0.92 & 0.65 & 0.71 & 0.76 & & 0.95 & 0.71 & 0.82 & 0.84 & & 0.99 & 0.86 & 0.93 & 0.93 \\
\textsc{qp} & & 0.94 & 0.91 & 0.90 & 0.93 & & 0.96 & 0.95 & 0.96 & 0.95 & & 0.99 & 1.00 & 0.99 & 1.00 \\
\end{tabular}
\end{table}
\begin{figure}[t]
\centering
\includegraphics[width=\textwidth]{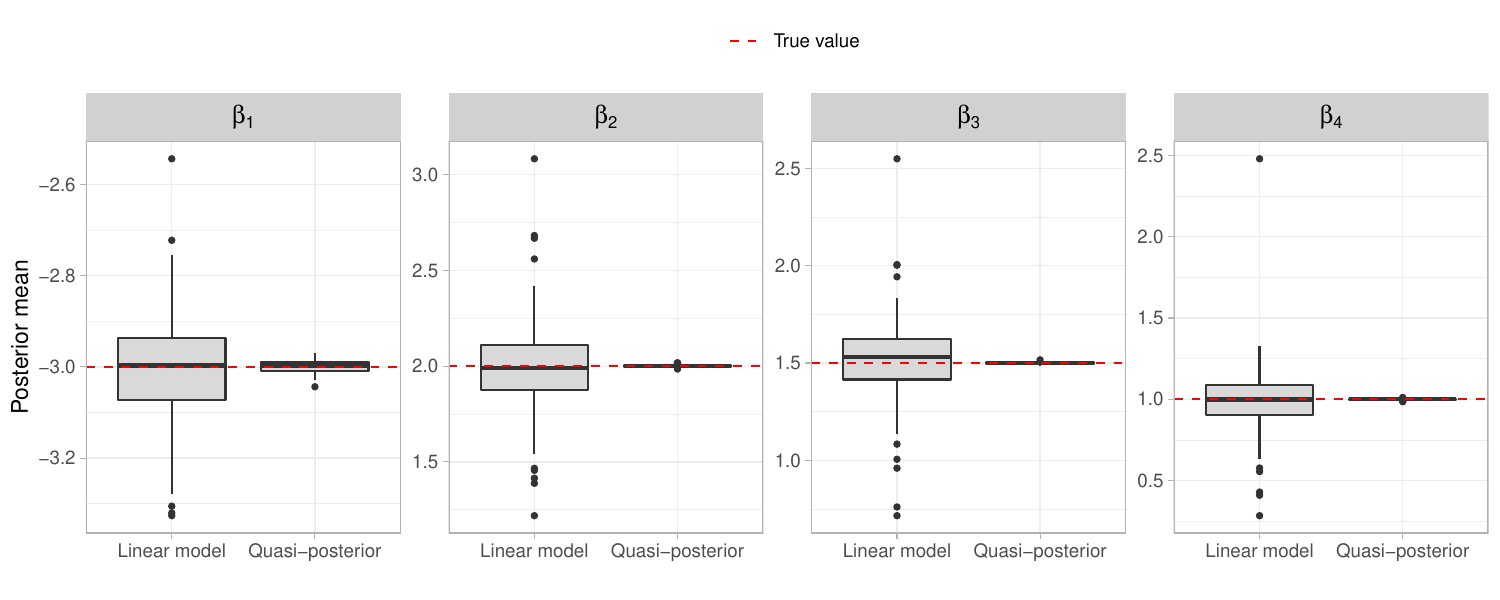}
\caption{\small Heteroscedastic continuous data example. Distribution of posterior mean of $\bmbeta$ obtained with Gaussian linear model and quasi-posterior.}
\label{fig:het_figure}
\end{figure}

The log-quasi-likelihood function contribution for the $i$th observation is $${\ell_\textsc{q}(\bmbeta;y_i,\x_i,\psi) = \psi^{-1}(\x_i^\top\bmbeta-y_i+1)\exp\{-\x_i^\top\bmbeta\}},$$ which leads to the quasi-posterior presented in Example \ref{ex:het_qp}.
In both the approaches we put an uninformative prior distribution on $\bmbeta$, and for the normal linear model an uninformative prior is put on the variance.
For the quasi-posterior, coherently with Section \ref{sec:connection_llb}, we estimate $\psi$ through the method of moments estimator \eqref{eqn:mom_psi}, where $\hbmbeta$ is the maximum quasi-likelihood estimate.
To approximately sample from both the posterior distributions we use a Hamiltonian Monte Carlo (HMC) algorithm provided by the Stan software \citep{stan}.
For each simulation, we run 3 chains of length 3,000 iterations discarding the first 1,000, for a total of 6,000 samples considered for inferential evaluations.

We assess performance by comparing accuracy of the 
estimated posterior means of $\bmbeta$ and frequentist coverage of credible intervals.
The distributions of the posterior means across the 100 simulations are reported in Figure \ref{fig:het_figure}.
Table \ref{tab:het_sim_covergage} shows empirical coverage of credible intervals.
The quasi-posterior means outperformed the posterior means under the Gaussian linear model.
Misspecification of the variance function in the normal linear model analysis leads to a loss of efficiency of the posterior means and poor frequentist coverage of credible intervals.
In contrast, even with a modest sample size, the quasi-posterior is able to produce well calibrated credible intervals.

\subsection{Simulation example: overdispersed count data}

It is very common for count data to exhibit over-dispersion relative to what simple models assume. To show effectiveness of quasi-posteriors in this context, we provide a simulation study comparing our approach with Poisson and negative binomial \textsc{glm}s, with the latter representing the 
usual model-based solution to overdispersion.
In particular, we consider the setting of Example \ref{ex:quasi_poisson}.

\begin{figure}[t]
\centering
\includegraphics[width=\textwidth]{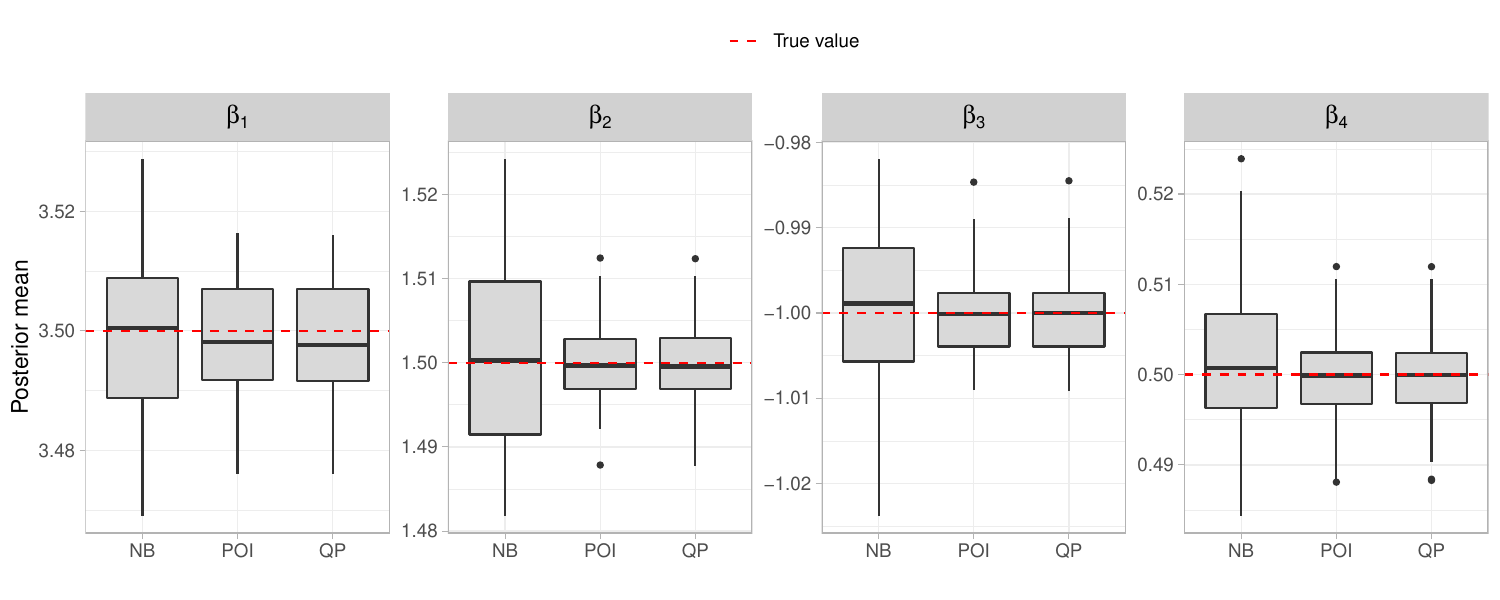}
\caption{\small Overdispersed count data example. Distribution of posterior mean of $\bmbeta$ obtained with negative binomial model (\textsc{nb}), Poisson model (\textsc{poi}) and quasi-posterior (\textsc{qp}).}
\label{fig:counts_figure}
\end{figure}

We generate $S = 100$ synthetic datasets of sample size $n = 1,000$.
In each simulation, the vector of observed responses is generated in two steps:
i) for $i=1,\ldots,n$, we draw $\Tilde{y}_i$ from $\mathrm{Ga}(\psi_0^{-1}\mu_{i,0}, \psi_0^{-1})$, where $\psi_0=3.5$, $\mu_{i,0} = \exp\{x_i^\top\bmbeta_0\}$, $\bmbeta_0 = (3.5,1.5,-1,0.5)$ and $\x_i$ are sampled from independent standard normal distributions;
ii) the corresponding observed response $y_i$ is obtained rounding $\Tilde{y}_i$ to the closest integer value.
As shown by \cite{tricker1984},  the effect of rounding on the mean of the gamma distribution is negligible and on the variance is very mild.
Thus, the $i$th observed response is drawn from a data generator that has mean $\mu_{i,0}$ and variance approximately equal to $\psi_0\mu_{i,0}$.
For the quasi-posterior, the dispersion parameter $\psi$ is estimated using \eqref{eqn:mom_psi} where in this case $\hbmbeta$ coincides with the maximum likelihood estimate.
Inference on posteriors and quasi-posteriors is carried out with an HMC algorithm using the Stan software.
For each simulation, we run 3 chains of length 1,000 iterations discarding the first 500, for a total of 1,500 samples considered for inferential evaluations.
Figure \ref{fig:counts_figure} compares the distributions of the posterior mean for $\bmbeta$ for the three models across the $S$ simulations, while Table \ref{tab:counts_sim_covergage} reports coverage of the corresponding $0.95$ credible intervals.

Poisson model and quasi-posterior have similar results in terms of efficiency. However, coverage differs substantially, with the Poisson model having frequentist coverage much lower than the nominal level.
As discussed in Example \ref{ex:quasi_poisson}, quasi-posterior \eqref{eqn:quasi_poisson_posterior} is equivalent to the power posterior obtained raising the Poisson likelihood to $\psi^{-1}$.
Poisson model and quasi-posterior express the relation between $\bmbeta$ and $\y$ with the same structure; consequently it is reasonable to expect that the two posteriors have similar means. However, the posterior concentration varies substantially.
The misspecified Poisson model produces smaller credible intervals, which tend to exclude the true parameter value if the posterior mean is not close to $\bmbeta_0$.
Conversely, the quasi-posterior is less concentrated with wider 
credible intervals.
As expected, the method of moments estimator is effective, 
leading to credible intervals that are well calibrated in their frequentist coverage.
This agrees with the asymptotic results of Section \ref{sec:coverage}.

The comparison between negative binomial and quasi-posterior settings offers additional interesting insights.
In terms of coverage calibration, both the models show good performance since they take into account the dispersion of the observations, although in two different manners: the first treats the dispersion as a model parameter and puts a prior on it, while the second uses its preliminary estimate to regulate the scale of the information coming from the data within a generalized Bayes approach.
However, the negative binomial \textsc{glm} is misspecified, leading to the loss of efficiency shown in Figure \ref{fig:counts_figure}.
The quasi-posteriors are substantially more concentrated than the negative binomial model posteriors, with narrower credible intervals that maintain well calibrated coverage.

\begin{table}[t]
\centering
\caption{\small Overdispersed count data example. Frequentist coverage of credible intervals with nominal level $\rho \in \{0.90, 0.95, 0.99\}$ for negative binomial model (\textsc{nb}), Poisson model (\textsc{poi}) and quasi-posterior (\textsc{qp}).}
\label{tab:counts_sim_covergage}
\vspace{0.2cm}
\footnotesize
\begin{tabular}{l c cccc c cccc c cccc}
 & & \multicolumn{4}{c}{$\rho = 0.90$} & & \multicolumn{4}{c}{$\rho = 0.95$} & & \multicolumn{4}{c}{$\rho = 0.99$}\\
 & & $\beta_1$ & $\beta_2$ & $\beta_3$ & $\beta_4$ & & $\beta_1$ & $\beta_2$ & $\beta_3$ & $\beta_4$ & & $\beta_1$ & $\beta_2$ & $\beta_3$ & $\beta_4$ \\
\textsc{nb} & & 0.79 & 0.84 & 0.92 & 0.94 & & 0.88 & 0.95 & 0.99 & 0.97 & & 0.91 & 1.00 & 0.99 & 0.99 \\
\textsc{poi} & & 0.59 & 0.62 & 0.63 & 0.64 & & 0.72 & 0.70 & 0.72 & 0.75 & & 0.89 & 0.80 & 0.86 & 0.85 \\
\textsc{qp} & & 0.94 & 0.94 & 0.92 & 0.88 & & 0.96 & 0.97 & 0.97 & 0.95 & & 1.00 & 1.00 & 0.99 & 0.99 \\
\end{tabular}
\end{table}

\subsection{Application: Rhynchosporium secalis data}
\label{sec:secalis}

\textit{Rhynchosporium secalis} is a fungus that can infect barley leaves causing leaf blotches.
\cite{wedderburn1974quasi} reported a study concerning the incidence of \textit{Rhynchosporium secalis} on $10$ different varieties of barley grown in $9$ sites, for a total of $n=90$ observations.
The response variable is a continuous proportion in $[0,1]$, representing the percentage of leaf area infected by the fungus.

\begin{figure}[t]
\centering
\includegraphics[width=0.75\textwidth]{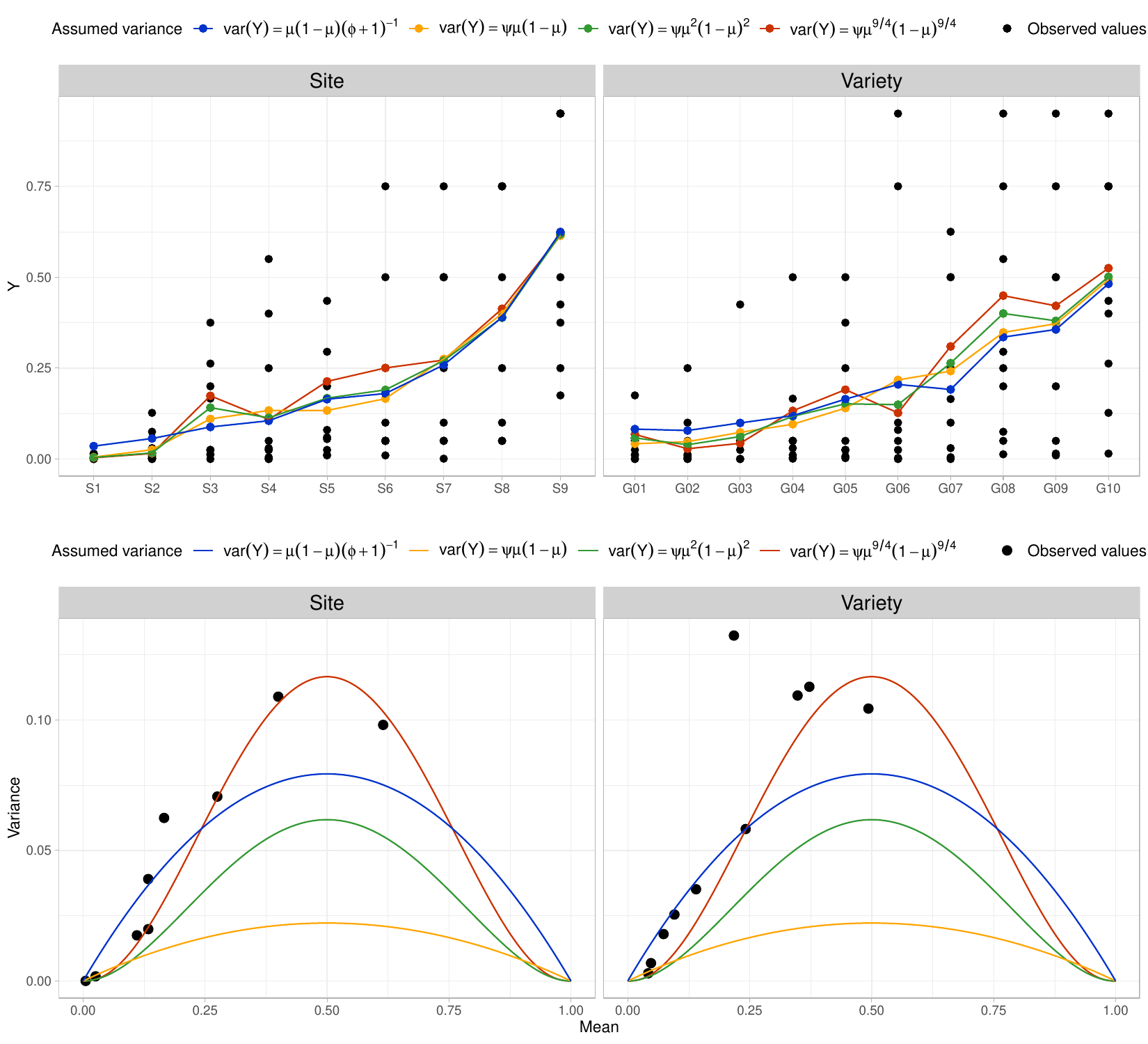}
\caption{\small Preliminary check of model appropriateness. 
The first row shows the comparison between the observed values and the means of the expected value obtained with \textsc{mle} for each site (top-left) and variety (top-right).
The second row shows the comparison of the observed relation between mean and variance with the theoretical ones for each site (bottom-left) and for each variety (bottom-right). For the quasi-posteriors, $\psi$ is estimated through the method of moments, while for beta regression, $\phi$ is estimated using the \textsc{mle}.
Sites and varieties are sorted in ascending order with respect to the corresponding mean of the response.}
\label{fig:secalis_fig}
\end{figure}

We first considered a full likelihood specification using beta regression, assuming a beta distribution with logit link function for the proportion of leaf area infected. Based on fitting such a model, we find that the variance function is sufficient for small values of the mean but as the mean increases there is underestimation of variance. In addition, small values of the mean seem to be overestimated.

This motivated us to consider quasi-likelihood. A first attempt used a quasi-binomial 
model with logit link function and $\var(Y_i) = \psi\mu_i(1-\mu_i)$, $i=1,\ldots,n$.
Results of preliminary model checking are shown in Figure \ref{fig:secalis_fig}.
The quasi-binomial model exhibits good performance in fitting the mean. However, the choice of variance function seems poor, as there is clear underestimation of the relationship between the mean and variance. In attempting to choose a better variance function, we considered the 
proposal of \cite{wedderburn1974quasi} to set $\var(Y_i) = \psi\mu_i^2(1-\mu_i)^2$. Unfortunately, the variance estimates are similar to the beta model, while the mean estimates are similar to the quasi-binomial.

An appealing aspect of quasi-likelihood is the great flexibility in choosing the variance function to better fit data having complex relationships. 
For these data, we find that a good choice for the variance is $\var(Y_i) = \psi\mu_i^{9/4}(1-\mu_i)^{9/4}$, as shown in Figure \ref{fig:secalis_fig}.
In this case both the estimated mean and variance provide a good fit to the observed data.

\begin{figure}[t]
\centering
\includegraphics[width=0.95\textwidth]{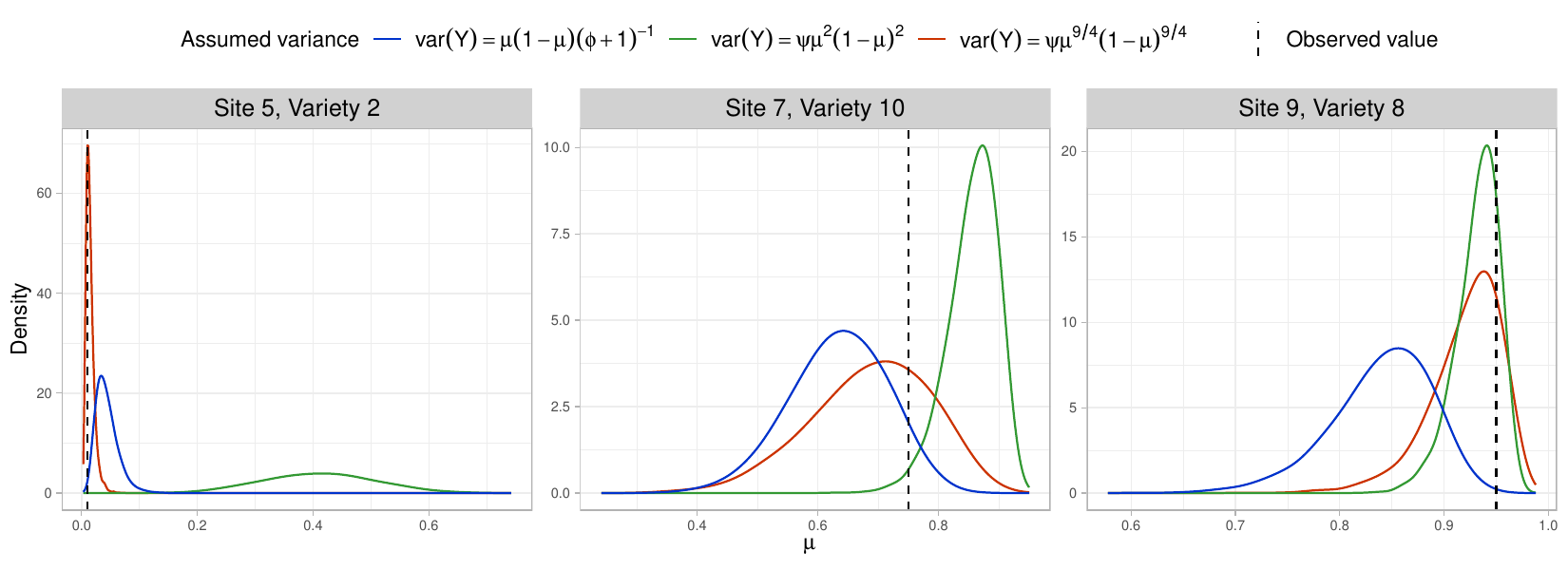}
\caption{\small Comparison of the posterior expectation for $\mu$ in three illustrative combinations of site and variety with small, central and large observed values, respectively.}
\label{fig:secalis_prev}
\end{figure}

We conduct detailed comparisons of Bayesian implementations of beta regression and quasi-posteriors assuming a logistic regression for the mean and variance functions that are such that either 
$\var(Y_i) = \psi\mu_i^2(1-\mu_i)^2$ or 
$\var(Y_i) = \psi\mu_i^{9/4}(1-\mu_i)^{9/4}.$
We put a weakly informative prior on $\bmbeta$ and use HMC and random walk Metropolis Hastings algorithms for sampling from the posterior and the quasi-posteriors, respectively.
The estimates of posterior mean for $\bmbeta$ differ across the approaches.
To assess goodness-of-fit, we check whether the 0.95 credible intervals for each $\mu_i$ contain the corresponding observed responses $y_i$.
For the quasi-posterior with Wedderburn variance function, 
the proportion of intervals containing the observed response is only $0.078.$
This proportion is $0.456$ for beta regression and $0.656$ for the proposed quasi-posterior, suggesting significant improvements in fit for the proposed approach.
Comparisons for three combinations of site and variety are illustrated in Figure \ref{fig:secalis_prev}.
Wedderburn quasi-posterior tends to overestimate the mean,
particularly for observations not close to one.
The proposed quasi-posterior works better than beta regression in each variety and in sites with either small or large mean.

\subsection{Application: willow warbler abundance in Finland}
\label{sec:birds}

As an additional application, we analyze data from an ecological monitoring study of birds focusing on 
occurrence rates of the willow warbler (\textit{Phylloscopus trochilus}) in Finland \citep{lindstrom_2015_birds} from 2006 to 2008.
The dataset consists of $n=244$ observations collected in $J=171$ different sites.
For each observation, the data include a set of environmental covariates that provide infomation about the habitat type (classified as either broadleaved forest, coniferous forest, open habitat, urban habitat, or wetlands) and the mean temperature registered in April-May.

The willow warbler is a long distance migrant passerine and is common in Finland certain times of the year.
After spending the winters in sub-Saharan Africa, this species arrives in the Fennoscandia (Baltic Shield) region in late April or beginning of May for the breeding season.
For each year, occurrences were surveyed across all sites between the end of May and beginning of July, which is after migration ends.
Willow warblers arriving in Finland have substantially different migratory routes than those breeding in Norway and southern Sweden \citep{hedenstrom1987migration}.
This species thrives in deciduous forests, but has adapted to living in heavily developed areas \citep{cramp1992handbook}.
Willow warbler abundance varies substantially across Finland, and our focus is on assessing to what extent this variation is explained by temperature and habitat features.

Standard practice in modeling of bird species abundance data focuses on using Poisson \textsc{glm}s that include site-specific random effects in addition to environmental covariates \citep{ovaskainen2020}.
However, bird occurrence counts tend to be overdispersed and zero-inflated relative to the Poisson. 
For the willow warbler data, the estimated dispersion is ${\Hat{\psi} = 6.57}$, suggesting substantial overdispersion relative to the Poisson. \cite{stoklosa2022overview} advocates for the general use of negative binomial models for species abundance in ecology. Hence, we focus on a
negative binomial hierarchical regression model as a competitor to our quasi-posterior approach. 
As zero counts are rare in these data ($0.4\%$ of the observations), zero-inflated and hurdle models do not seem appropriate.
We assume the following mean and variance functions in constructing our quasi-posterior:
\begin{equation*}
    \E(Y_{ij}) = \mu_{ij}, \quad 
    \var(Y_{ij}) = \psi \mu_{ij}, \quad
    \mu_{ij} = \exp\{\x_{ij}^\top\bmbeta + \delta_j\}, \quad
    \delta_j \sim \mathcal{N}(0, \sigma),
\end{equation*}
for $i = 1,\ldots,n_j$, $j=1,\ldots,J$, $\sum_{j=1}^Jn_j = n$. We focus on the log link, as it leads to easy interpretation of regression coefficients and is standard in ecological abundance modeling.

The covariate vector $\x_{ij}$ 
includes temperature, habitat, and year. Specifically, the temperature variable corresponds to the average temperature (in degrees Celsius) recorded during April and May in each site for the corresponding year. The average temperature during these months may impact bird preferences for breeding sites and movement patterns. The different habitats mentioned above are included as dummy indicators, with broadleaved forest considered as the baseline as the primary native habitat of this species.
Average spring temperature and habitat for each location are reported in Appendix \ref{apx:birds_cov}.
Finally, the survey year is included to account for possible annual effects.
This is encoded in the covariates into two dummy variables, taking 2006 as baseline.

\begin{table}[t]
\centering
\caption{\small Posterior mean and 0.95 credible intervals for $\bmbeta$ components and $\sigma$.
Parameters in italics refer to habitat categories (\textit{broadleaved forest} is taken as baseline), while \textit{2006} is the baseline for year components.}
\label{tab:birds}
\vspace{0.2cm}
\footnotesize
\begin{tabular}{l c rl c rl c rl}
& & \multicolumn{2}{c}{\textsc{poi}} & & \multicolumn{2}{c}{\textsc{nb}} & & \multicolumn{2}{c}{\textsc{qp}}\\
Intercept & & $4.10$ & [$3.80$, $4.38$] & & $3.71$ & [$3.43$, $3.98$] & & $3.74$ & [$3.47$, $3.99$] \\
Temperature & & $-0.07$ & [$-0.10$, $-0.03$] & & $0.03$ & [$-0.01$, $0.07$] & & $0.02$ & [$-0.01$, $0.05$] \\
Year \\
\:\:\textit{2007} & & $0.28$ & [$0.21$, $0.35$] & & $0.30$ & [$0.11$, $0.49$] & & $0.27$ & [$0.11$, $0.41$] \\
\:\:\textit{2008} & & $-0.03$ & [$-0.10$, $0.05$] & & $0.03$ & [$-0.14$, $0.20$] & & $0.00$ & [$-0.15$, $0.14$] \\
Habitat \\
\:\:\textit{Coniferous f.} & & $-0.26$ & [$-0.50$, $-0.01$] & & $-0.25$ & [$-0.47$, $-0.03$] & & $-0.24$ & [$-0.45$, $-0.06$] \\
\:\:\textit{Open habitat} & & $-0.74$ & [$-1.07$, $-0.43$] & & $-0.44$ & [$-0.72$, $-0.13$] & & $-0.42$ & [$-0.69$, $-0.17$] \\
\:\:\textit{Urban habitat} & & $-0.54$ & [$-0.88$, $-0.19$] & & $-0.77$ & [$-1.09$, $-0.46$] & & $-0.73$ & [$-1.03$, $-0.42$] \\
\:\:\textit{Wetlands} & & $-0.37$ & [$-0.79$, $0.03$] & & $-0.16$ & [$-0.54$, $0.22$] & & $-0.18$ & [$-0.52$, $0.17$] \\
Site \\
\:\:$\sigma$ & & $0.34$ & [$0.25, 0.45]$ & & $0.07$ & $(0.00, 0.14]$ & & $0.34$ & $[0.27, 0.42]$ \\
\end{tabular}
\end{table}

The inclusion of random intercepts $\bm{\delta}=(\delta_1,\ldots,\delta_J)$ 
accounts for the hierarchical structure of the data and heterogeneity among the locations in willow warbler occurrence rates; for example, this heterogeneity may be due to differences in habitat that are not captured by measured covariates.
The common Gaussian prior centered at zero shrinks the site-specific random intercepts toward the population mean.

We compare our quasi-posterior with Poisson and negative binomial regression models having an identical mean structure to ours, including site-specific random effects.
We put weakly informative priors on $\bmbeta$ and $\sigma$. 
For each model, we sample from the posterior distribution using an HMC algorithm in Stan running 4 chains of length 2,000 iterations discarding the first 500, for a total of 6,000 samples considered for inferential evaluations.
To assess goodness-of-fit, we compare the in-sample standardized mean squared error (sMSE) of the three models computed using Pearson residuals, which remove the effect
of assumed heteroscedasticity \citep{agresti2015foundations}.
As expected, the Poisson model has the highest sMSE ($1.72$) since it underestimates the dispersion, while the quasi-posterior and negative binomial model have substantially better performances since both account for overdispersion, with a better fit for the quasi-posterior ($0.59$ sMSE vs $0.65$ for the negative binomial).
For this reason, we focus on the inferential conclusions of these two.

\begin{figure}[t]
\centering
\includegraphics[width=0.8\textwidth]{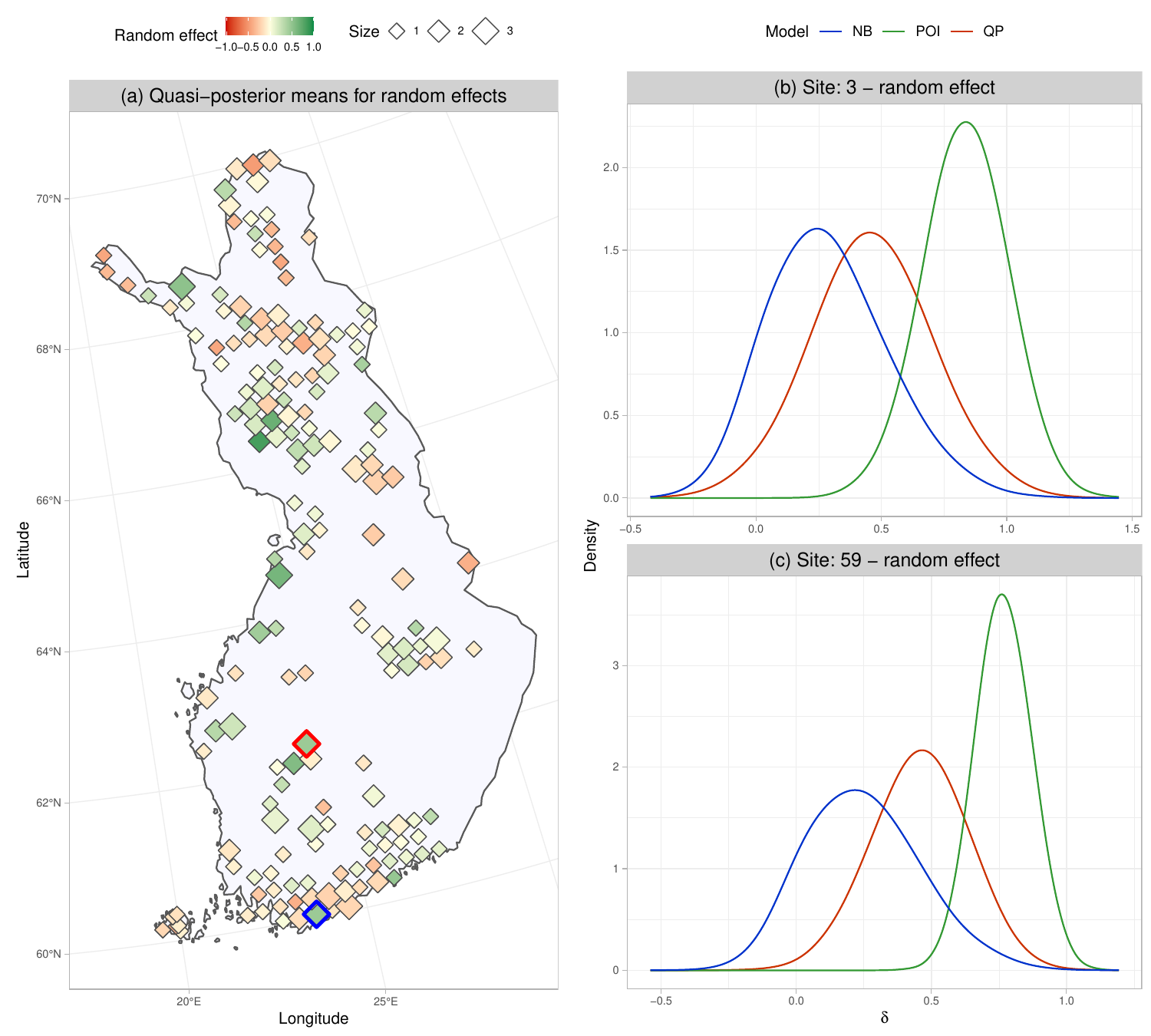}
\caption{\small 
Panel (a) represents the quasi-posterior means of site-specific random effects and their geographical location in Finland.
Site 3 is highlighted in blue and its posteriors distributions are represented in panel (b).
Site 59 is highlighted in red and its posteriors distributions are represented in panel (c).}
\label{fig:birds}
\end{figure}

Posterior means and 0.95 credible intervals for $\bmbeta$ are reported in Table \ref{tab:birds}.
As expected, the negative binomial model and quasi-posterior provide similar estimates for the regression coefficients, and their posterior distributions have higher variance than those under the Poisson model.
The results suggest that 2007 was favorable for willow warbler abundance in Finland compared to 2006 and 2008.
Causes of the differences in abundance across years may be due to unmeasured weather covariates.
Willow warbler abundance is lower in coniferous forests, open and urban habitats compared to broadleaved forests. These conclusions are in line with literature noting that willow warblers prefer to feed in deciduous forests
\citep{virkkala1991spatial} and that in urban areas there is a reduction of the number of \textit{Phylloscopus trochilus} males due to traffic \citep{reijnen1991effect, reijnen1994effects}. 
Moreover, \textit{Phylloscopus trochilus} diffusion does not significantly vary according to the average spring temperature across the locations. 
This indicates that the temperature the birds encounter in Finland after their spring migration does not affect their spreading over the whole territory, which is influenced mainly by the habitat.

The posterior distribution of the random intercepts helps identify differences in willow warbler abundance across locations that are not explained by our observed habitat and temperature covariates.
In interpreting these results, we focus on our quasi-posterior for brevity.
As shown in Figure \ref{fig:birds}a,
sites having unusually low willow warbler abundance are mainly located in southern Finland and upper Lapland, while 
sites with unusually high abundance concentrate 
along the Bothnian Bay coast and in lower Lapland.
It is likely that these spatial patterns reflect additional more nuanced habitat features not captured in the coarse habitat classification provided in the data for each site. In future work it will be interesting to look more into these patterns to infer these latent habitat factors to learn more about what drives bird population dynamics, and expand habitat classification in future studies.

Probing the posterior on the random effects
allows identification of interesting locations.
For example, Figure \ref{fig:birds}b and \ref{fig:birds}c show two sites with unexpected high abundances - one in urban Helsinki and another in a central Finland 
coniferous forest.
The former may represent a success story in terms of providing a good habitat within a city environment. This success story may inform future policies for protecting green areas in cities. The later would be interesting to follow-up to assess whether there is something special about that coniferous forest causing high abundance of a deciduous forest species.

\section{Discussion}

Our proposed quasi-posteriors provide a generalized Bayes solution to overcome misspecification problems in Bayesian \textsc{glm}s. Quasi-posteriors additionally have interesting links to c-posteriors, giving new insights into coarsening. We have demonstrated that our quasi-posteriors are highly versatile, allowing one to focus on mean and variance functions in model specification, simplifying the search for an appropriate model for data with challenging heteroscedasticity. An interesting application is to count response data, providing notable improvements over not only Poisson regression but also negative binomial-based \textsc{glm}s. We also see substantial improvements for continuous response data having variance changing with covariates. Addition of random effects is straightforward, making our method convenient for many different applications. The code to replicate our analyses is available online in the repository \url{https://github.com/davideagno/Quasi-posteriors}.

Quasi-posteriors are designed for inference on regression coefficients. For simplicity and ease in general implementation, we use a method of moments estimator for the dispersion parameter $\psi$, following standard practice in the frequentist literature. An interesting area of future research is to develop approaches for uncertainty quantification for $\psi$. Potentially a prior can be chosen for $\psi$ and a Bayesian approach can be developed for inference, building on methods in \cite{bissiri2016general}.
However, there are some associated subtleties that would need to be addressed.

An additional interesting area for future research is the development of formal model selection procedures. We illustrated one heuristic strategy in our analysis in Section \ref{sec:secalis}; in particular, in this case we considered a variety of different variance functions and selected the one with the best goodness-of-fit to the data. As alternative model performance criteria for use in model comparisons, one could rely on 
leave-one-out cross-validation  and marginal quasi-likelihood as in \cite{fong2020marginal}, or adaptations of deviance information criteria and related ideas \citep{gelman2013bayesian}. This can be applied to selection of both the mean and variance functions. Interesting next steps include theoretical support for such model selection strategies, including frequentist asymptotic theory and a Bayesian decision theoretic justification.

\section*{Funding}

This research was partially supported by the Lifeplan project funded by the European Research Council under the European Union’s Horizon 2020
research and innovation programme (grant agreement No 856506). Additional funding came from the United States National Institutes of Health R01-ES028804 and R01-ES035625.


\appendix
\section{Proofs}

\subsection{Proof of Theorem \ref{th:beta_opt_zero}}\label{proof_one}

The inferential target for Gibbs posterior is
\begin{equation*}
    \bmbeta_\textsc{opt} = \arg\max_{\bmbeta}\int\ell_\textsc{q}(\bmbeta,\y,\X,\psi)F_0(\dd \y \mid \X) = \arg\max_{\bmbeta}\mathcal{L}(\bmbeta; \bmbeta_0, \X, \psi),
\end{equation*}
where $F_0(\dd \y \mid \X)$ denotes the probability measure corresponding to the density $f_0(\y\mid\X) = f(\y\mid\X,\bmbeta_0,\psi_0) = \prod_{i=1}^n f(y_i\mid\x_i,\bmbeta_0,\psi_0)$, since $Y_i\mid\x_i$ are independent by assumption, and having defined
\begin{equation*}
    \mathcal{L}(\bmbeta; \bmbeta_0, \X, \psi)  =
    \sum_{i=1}^n \mathcal{L}(\bmbeta; \bmbeta_0, \x_i, \psi)=
    \sum_{i=1}^n \int_\mathcal{Y}\int_a^{\mu_i}\frac{y_i-t}{\psi V(t)}\dd t F_0(\dd y_i \mid \x_i),
\end{equation*}
where $\mu_i = \mu_i(\bmbeta) = g^{-1}(\x_i^\top\bmbeta) \in \mathcal{M} \subseteq \mathds{R}$. By Fubini-Tonelli theorem, the integral
\begin{equation*}
    \int_{\mathcal{Y}\times[a,\mu_i]} \left|\frac{y_i-t}{\psi V(t)}\right|\dd tF_0(\dd y_i \mid \x_i) =
    \int_\mathcal{Y}\int_a^{\mu_i} \left|\frac{y_i-t}{\psi V(t)}\right|\dd tF_0(\dd y_i \mid \x_i) =
    \int_a^{\mu_i}\int_\mathcal{Y} \left|\frac{y_i-t}{\psi V(t)}\right|F_0(\dd y_i \mid \x_i)\dd t
\end{equation*}
is well defined for any $i=1,\ldots,n$, albeit potentially infinite.
By assumption, the variance function $\psi V(t)$ is strictly positive and $\int|y_i|F_0(\dd y_i \mid \x_i) \le \left(\int y_i^2 F_0(\dd y_i \mid \x_i)\right)^{1/2} = c_i < \infty$ because the second moment is finite for all $i=1,\ldots,n$. Then, as an application of the triangular inequality, we obtain the following chain of inequalities
\begin{equation*}
    \int_a^{\mu_i}\int_\mathcal{Y} \left|\frac{y_i-t}{\psi V(t)}\right|F_0(\dd y_i \mid \x_i)\dd t 
    \le\int_a^{\mu_i} \frac{1}{\psi V(t)}\left(\int_\mathcal{Y}|y_i|F_0(\dd y_i \mid \x_i) + |t| \right)\dd t
    \le\int_a^{\mu_i} \frac{c_i-|t|}{\psi V(t)}\dd t
    < \infty,
\end{equation*}
where the last inequality follows because the function $(c_i-|t|)\psi^{-1}V(t)^{-1}$ is continuous and bounded for any $t\in[a,\mu_i]$ and $i=1,\ldots,n$.
Thus, the function $(y_i-t)\psi^{-1}V(t)^{-1}$ is integrable in the product space $\mathcal{Y}\times[a,\mu_i]$ with respect to the product measure and the order of integration can be interchanged, giving
\begin{equation*}
    \mathcal{L}(\bmbeta; \bmbeta_0, \X, \psi) =
    \sum_{i=1}^n \int_a^{\mu_i}\int_{\mathcal{Y}}\frac{y_i-t}{\psi V(t)}F_0(\dd y_i \mid \x_i)\dd t
    = \sum_{i=1}^n \int_a^{\mu_i} \frac{g^{-1}(\x_i^\top\bmbeta_0) - t}{\psi V(t)} \dd t,
\end{equation*}
where $g^{-1}(\x_i^\top\bmbeta_0) = \int_{\mathcal{Y}} y_i F_0(\dd y_i \mid \x_i)$.
The gradient vector
\begin{equation*}
    \nabla\mathcal{L}(\bmbeta; \bmbeta_0, \X, \psi)
    = \frac{\partial}{\partial\bmbeta} \sum_{i=1}^n \int_a^{\mu_i} \frac{g^{-1}(\x_i^\top\bmbeta_0) - t}{\psi V(t)} \dd t
    = \sum_{i=1}^n \frac{g^{-1}(\x_i^\top\bmbeta_0) - g^{-1}(\x_i^\top\bmbeta)}{\psi V(\mu_i)g'(\mu_i)}\x_i
\end{equation*}
is zero if $\bmbeta = \bmbeta_0$. The vector $\bmbeta_0$ is a maximum because the second derivative $\nabla^2\mathcal{L}(\bmbeta; \bmbeta_0, \X, \psi)$, evaluated in $\bmbeta_0$, is negative definite. In fact, the expected information defined in \eqref{eqn:information_ql} evaluated in $\bmbeta_0$ equals to
$$
I(\bmbeta_0)= - \nabla^2\mathcal{L}(\bmbeta; \bmbeta_0, \X, \psi) |_{\bmbeta = \bmbeta_0} = \psi^{-1}\bm{X}^\top \Tilde{\bm{D}}\bm{X},
$$ 
where $\Tilde{\bm{D}} = \text{diag}(\Tilde{d}_1,\ldots,\Tilde{d}_n)$, $\Tilde{d}_i = \big[ V(\mu_i(\bmbeta_0)) g'(\mu_i(\bmbeta_0))^2 \big]^{-1}$ for $i=1,\ldots,n$,
and is positive definite because the design matrix $\bm{X}$ has full rank for any $n \ge p$ and the entries $\Tilde{d}_i > 0$ are strictly positive by assumption.

The value $\bmbeta_0$ is the only solution in many concrete scenarios. For instance, suppose the identity $g'(\mu) = 1 / V(\mu)$ holds, corresponding to a quasi-likelihood with canonical link such as overdispersed Poisson regression with log link and logistic regression; then the integrated loss $\mathcal{L}(\bmbeta; \bmbeta_0, \X, \psi)$ is concave, because $I(\bmbeta)= - \nabla^2\mathcal{L}(\bmbeta; \bmbeta_0, \X, \psi)$ for all values of $\bmbeta$. In this case, $\bmbeta_0$ is the unique solution.

\subsection{Proof of Proposition \ref{prop:cposterior}}

The power posterior approximation of the corresponding c-posterior based on \eqref{eqn:exp_fam} is
\begin{equation}
    p_\alpha(\bmbeta\mid\y,\X) \propto p(\bmbeta) \prod_{i=1}^n \exp\left\{ \frac{\alpha}{\alpha+n} \Big [\theta(\mu_i)y_i - b(\theta(\mu_i))\Big]\right\}.
    \label{eqn:power_posterior_exp_fam}
\end{equation}
By assumption, the second order conditions \eqref{eqn:second_order} are such that $\var(Y_i) = \psi V(\mu_i)$, $i=1,\ldots,n$, with
\begin{equation*}
    V(\mu_i) = \frac{\partial^2}{\partial\theta^2} b(\theta) \at[\bigg]{\theta=\theta(\mu_i)},
\end{equation*}
and, by Theorem \ref{th:wedderburn}, the corresponding quasi-posterior is
\begin{equation*}
    p_\textsc{q}(\bmbeta\mid\y,\X,\psi) \propto p(\bmbeta)\prod_{i=1}^n \exp\left\{ \frac{1}{\psi} \Big [\theta(\mu_i)y_i - b(\theta(\mu_i))\Big]\right\},
\end{equation*}
that is equivalent to \eqref{eqn:power_posterior_exp_fam} for $\psi = (\alpha+n)/\alpha$.

\subsection{Proof of Theorem \ref{th:asymptotic} and regularity conditions}
\label{proof_asymptotics}

The result follows from the general theory developed in Theorems 4 and 5 of \cite{miller2021asymptotic}, in which are presented abstract sufficient conditions for which a generalized posterior converges in total variation to a Gaussian distribution. We review these conditions and show how they can be specialized in the context of quasi posteriors. Let $B \subseteq \mathds{R}^p$ be an open and convex set such that $\bmbeta_0 \in B$. In the first place,
 we need to ensure that the function $\ell_\textsc{q}(\bmbeta;\y,\X,\psi)$ is well-behaved, namely that $\ell_\textsc{q}(\bmbeta;\y,\X,\psi)$ has continuous third derivatives on $B$ and that the third derivative is uniformly bounded on $B$; these are very mild conditions that are satisfied for any reasonable choice of the link $g(\cdot)$ and variance $V(\cdot)$ functions.
 In the second place, it is required in \citet[][Theorem 5]{miller2021asymptotic} that the following quantity 
\begin{equation*}\label{eq:cond1}
\frac{1}{n}\ell_\textsc{q}(\bmbeta;\y,\X,\psi) = \frac{1}{n}\sum_{i=1}^n\ell_\textsc{q}(\bmbeta; y_i, \bm{x}_i, \psi)  = \frac{1}{n}  \sum_{i=1}^n \frac{1}{\psi}\int_{a}^{\mu_i(\bmbeta)} \frac{y_i - t}{V(t)} \dd t,
\end{equation*}
converges pointwise and almost surely to a well defined limit, say $\ell_\infty(\bmbeta)$, for any $\bmbeta \in B$, as $n \rightarrow \infty$. In quasi-posteriors, the contributions $\ell_\textsc{q}(\bmbeta; y_1, \bm{x}_1, \psi), \dots, \ell_\textsc{q}(\bmbeta; y_n, \bm{x}_n, \psi)$ are independent realizations, therefore extensions of the law of large numbers can be invoked. Note that each $\mathcal{L}(\bmbeta; \bmbeta_0, \x_i, \psi) = \mathds{E}(\ell_\textsc{q}(\bmbeta; y_i, \bm{x}_i, \psi))$ is finite, as already discussed in Section~\ref{proof_one}. Hence, sufficient conditions for pointwise convergence to $\ell_\infty(\bmbeta)$ are 
$$
\lim_{n\rightarrow \infty} \frac{1}{n}\sum_{i=1}^n\mathcal{L}(\bmbeta; \bmbeta_0, \x_i, \psi) = \ell_\infty(\bmbeta)< \infty, \qquad \sum_{n=1}^\infty \frac{\text{var} (\ell_\textsc{q}(\bmbeta; Y_n, \bm{x}_n, \psi))}{n^2} < \infty.
$$
as a consequence of Proposition 6.13, \citet[][Chapter 3]{Cinlar2011}. Both these conditions are very mild and essentially requires that the quasi-log-likelihood is well-defined at the limit. Thirdly, regularity conditions on the limiting function $\ell_\infty(\bmbeta)$ are needed. In particular, it is required that $i(\bmbeta_0) = - \nabla^2 \ell_\infty(\bmbeta_0)$ is a positive definite matrix. As shown in the proof of Appendix~\ref{proof_one}, $I(\bmbeta_0)= - \nabla^2\mathcal{L}(\bmbeta; \bmbeta_0, \X, \psi) |_{\bmbeta = \bmbeta_0} = \psi^{-1}\bm{X}^\top \Tilde{\bm{D}}\bm{X}$ is positive definite. Hence, recalling that $\ell_\infty(\bmbeta) = \lim_{n\rightarrow \infty} n^{-1}\sum_{i=1}^n\mathcal{L}(\bmbeta; \bmbeta_0, \x_i, \psi)$, we assume that $i(\bmbeta_0) =\lim_{n\rightarrow \infty} n^{-1} I(\bmbeta_0) = \lim_{n\rightarrow \infty} (n\psi)^{-1}\bm{X}^\top \Tilde{\bm{D}}\bm{X}$ is positive definite. Finally, either of the following two conditions must be satisfied:
\begin{enumerate}
\item[i)] $\ell_\infty(\bmbeta) < \ell_\infty(\bmbeta_0)$ for all $\bmbeta \in B_* \setminus \{\bmbeta_0\}$ and $\lim\sup_{n}\sup_{\bmbeta \in \mathds{R}^p \setminus B_*} \ell_\textsc{q}(\bmbeta;\y,\X,\psi) < \ell_\infty(\bmbeta_0)$ for some compact set $B_* \subseteq B$ with $\bmbeta_0$ in the interior of $B_*$;
\item[ii)] $\ell_\textsc{q}(\bmbeta;\y,\X,\psi)$ is concave and $\nabla\ell_\infty(\bmbeta_0) = 0$.
\end{enumerate}
The requirement (i) is a Wald-type condition and essentially it means that the function $\ell_\infty(\bmbeta)$ is maximized in $\bmbeta_0$. Condition (ii) is stronger and it implies (i), but is satisfied in many practical occasions. Indeed, paralleling the reasoning of the proof of Appendix~\ref{proof_one}, if the identity $g'(\mu) = 1 / V(\mu)$ holds, then the loss $\ell_\textsc{q}(\bmbeta;\y,\X,\psi)$ is concave, because we get $I(\bmbeta)= - \nabla^2\ell_\textsc{q}(\bmbeta;\y,\X,\psi)$, which is positive definite for all values of $\bmbeta$. Moreover, note that $\nabla \ell_\infty(\bmbeta_0) = \lim_{n\rightarrow \infty} n^{-1}\nabla \mathcal{L}(\bmbeta; \bmbeta_0, \X, \psi)|_{\bmbeta = \bmbeta_0} = 0$, as shown in Appendix~\ref{proof_one}.

\subsection{Proof of Theorem \ref{th:coverage}}
\label{proof_coverage}

The result follows from Theorem 8 in \cite{miller2021asymptotic}, that provides the sufficient conditions for which a generalized posterior exhibits correct frequentist coverage. In the first place, we require the convergence in distribution of $-\sqrt{n}(\hbmbeta-\bmbeta_0)$ to a random variable $Z$ as $n\rightarrow\infty$. In our case this is true because, under the regularity assumptions of Theorem~\ref{th:asymptotic}, we have $\sqrt{n}(\hbmbeta - \bmbeta_0) \xrightarrow{d} Z$, with $Z\sim N(0,i(\bmbeta_0)^{-1})$; see e.g. \citet{mccullagh1983quasi}. The second condition requires that the distribution $\pi_n$ of $\sqrt{n}(\bmbeta-\hbmbeta)$ converges in total variation to the law of $Z$, which is guaranteed by Theorem~\ref{th:asymptotic} under the regularity conditions presented in Appendix~\ref{proof_asymptotics}. The third condition in \citet{miller2021asymptotic} requires the pointwise convergence of the sequence of centered and rescaled credible sets $C_n = \{\sqrt{n}(\bmbeta-\hbmbeta)\in\R^p : \bmbeta\in S_n\}$ to a fixed set $C \subseteq \R^p$ with finite nonzero Lebesgue measure, that is, for all $\bm{u} \in \R^p$, $\mathds{1}(\bm{u} \in C_n) \rightarrow \mathds{1}(\bm{u} \in C)$ almost surely as $n\rightarrow\infty$. The validity of this requirement depends on the specific choice of the credible sets. As an example, fix $C = \{\bm{u} \in \R^p : \mathcal{N}(\bm{u}\mid 0, i(\bmbeta_0)^{-1}) \ge k_\rho\}$,
 where $\mathcal{N}(\cdot\mid 0, i(\bmbeta_0)^{-1})$ denotes the density function of $Z$,
and $k_\rho$ is such that $\int_C \mathcal{N}(\bm{u}\mid 0, i(\bmbeta_0)^{-1})\dd\bm{u} = \rho$.
Suppose $p_\textsc{q}(\bmbeta\mid\y,\X,\psi)$ is unimodal  and $S_n$ are highest posterior density credible sets. 
Then $C_n = \{\bm{u} \in \R^p : p_n(\bm{u}) \ge k_\rho \}$, where $p_n(\cdot)$ is the density function of $\sqrt{n}(\bmbeta-\hbmbeta)$.
Then,
\begin{align*}
        \mathds{1}(\bm{u}\in C_n) = 1 \quad & \text{if and only if}\quad
        p_n(\bm{u}) \ge k_\rho,\\
        \mathds{1}(\bm{u}\in C) = 1 \quad & \text{if and only if}\quad
        \mathcal{N}(\bm{u}\mid0,i(\bmbeta_0)^{-1}) \ge k_\rho.
\end{align*}
Therefore, for all $\bm{u}\in\R^p$, $\mathds{1}(\bm{u} \in C_n) \xrightarrow{a.s.} \mathds{1}(\bm{u} \in C)$ as $n\rightarrow \infty$, because 
\begin{equation*}
        \int_{\R^p} \left| p_n(\bm{u}) - \mathcal{N}(\bm{u}\mid0,i(\bmbeta_0)^{-1})\right|\dd\bm{u} \rightarrow 0
\end{equation*}
almost surely under $f_0$ by our Theorem~\ref{th:asymptotic}. The final condition in \citet{miller2021asymptotic} is automatically satisfied as $Z$ is absolutely continuous, being a multivariate Gaussian.

\subsection{Proof of Proposition \ref{prop:psi_lyddon}}

Recalling \eqref{eqn:loss_function}, we can write $\ell_\textsc{q}(\bmbeta,\y,\X,\psi) = \sum_{i=1}^n\ell_\textsc{q}(\bmbeta,y_i,\x_i,\psi) = - \psi^{-1}\sum_{i=1}^n\ell(\bmbeta;y_i,\x_i)$ with 
\begin{equation*}
\ell(\bmbeta;y_i,\x_i) = - \int_a^{\mu_i} \frac{y_i - t}{V(t)}\dd t,
\label{eqn:loss_ql}
\end{equation*}
where $\mu_i = \mu_i(\bmbeta) = g^{-1}(\x_i^\top\bmbeta) \in \mathcal{M} \subseteq \mathds{R}$.
Then
\begin{align*}
    J(\bmbeta) & = \int\nabla^2\ell(\bmbeta;\y,\X)F_0(\dd \y\mid\X) 
    = \sum_{i=1}^n\int\nabla^2\ell(\bmbeta;y_i,\x_i)F_0(\dd y_i\mid\x_i),\\
    H(\bmbeta) & = \int\nabla\ell(\bmbeta;\y,\X)\nabla\ell(\bmbeta;\y,\X)^\top F_0(\dd \y\mid\X) 
    = \sum_{i=1}^n\int\nabla\ell(\bmbeta;y_i,\x_i)\nabla\ell(\bmbeta;y_i,\x_i)^\top F_0(\dd y_i\mid\x_i).
\end{align*}
Recall that $\int y_i F_0(\dd y_i\mid\x_i) = g^{-1}(\x_i^\top\bmbeta_0) = \mu_i(\bmbeta_0)$ and $\int [y_i-\mu_i(\bmbeta_0)]^2 F_0(\dd y_i\mid\x_i) = \psi_0 V(\mu_i(\bmbeta_0))$. Hence, after some algebra \citep{wedderburn1974quasi}, we obtain
\begin{equation*}
    J(\bmbeta_0) = \X^\top\Tilde{\bm{D}}\X, \qquad  H(\bmbeta_0) = \psi_0 \X^\top\Tilde{\bm{D}}\X
    = \psi_0 J(\bmbeta_0).
\end{equation*}
Consequently,
\begin{align*}
    j(\bmbeta_0) & = \lim_{n\rightarrow\infty} n^{-1} \E\left[\nabla^2\ell(\bmbeta_0;\Y,\X)\right] = \lim_{n\rightarrow\infty} n^{-1}J(\bmbeta_0),\\
    h(\bmbeta_0) & = \lim_{n\rightarrow\infty} n^{-1}\E\left[\nabla\ell(\bmbeta_0;\Y,\X)\nabla\ell(\bmbeta_0;\Y,\X)^\top\right] = \lim_{n\rightarrow\infty} n^{-1}H(\bmbeta_0) = \psi_0 j(\bmbeta_0).
\end{align*}
Therefore, \eqref{eqn:psi_lyddon} can be simplified as
\begin{equation*}
    \psi_\textsc{llb} = 
    \frac{\text{tr}\{j(\bmbeta_0)\}}{\text{tr}\{j(\bmbeta_0)h(\bmbeta_0)^{-1}j(\bmbeta_0)\}} = \psi_0.
\end{equation*}

\clearpage
\section{Summary of environmental covariates of willow warblers abundance analysis}
\label{apx:birds_cov}

\begin{figure}[h]
\centering
\includegraphics[width=0.9\textwidth]{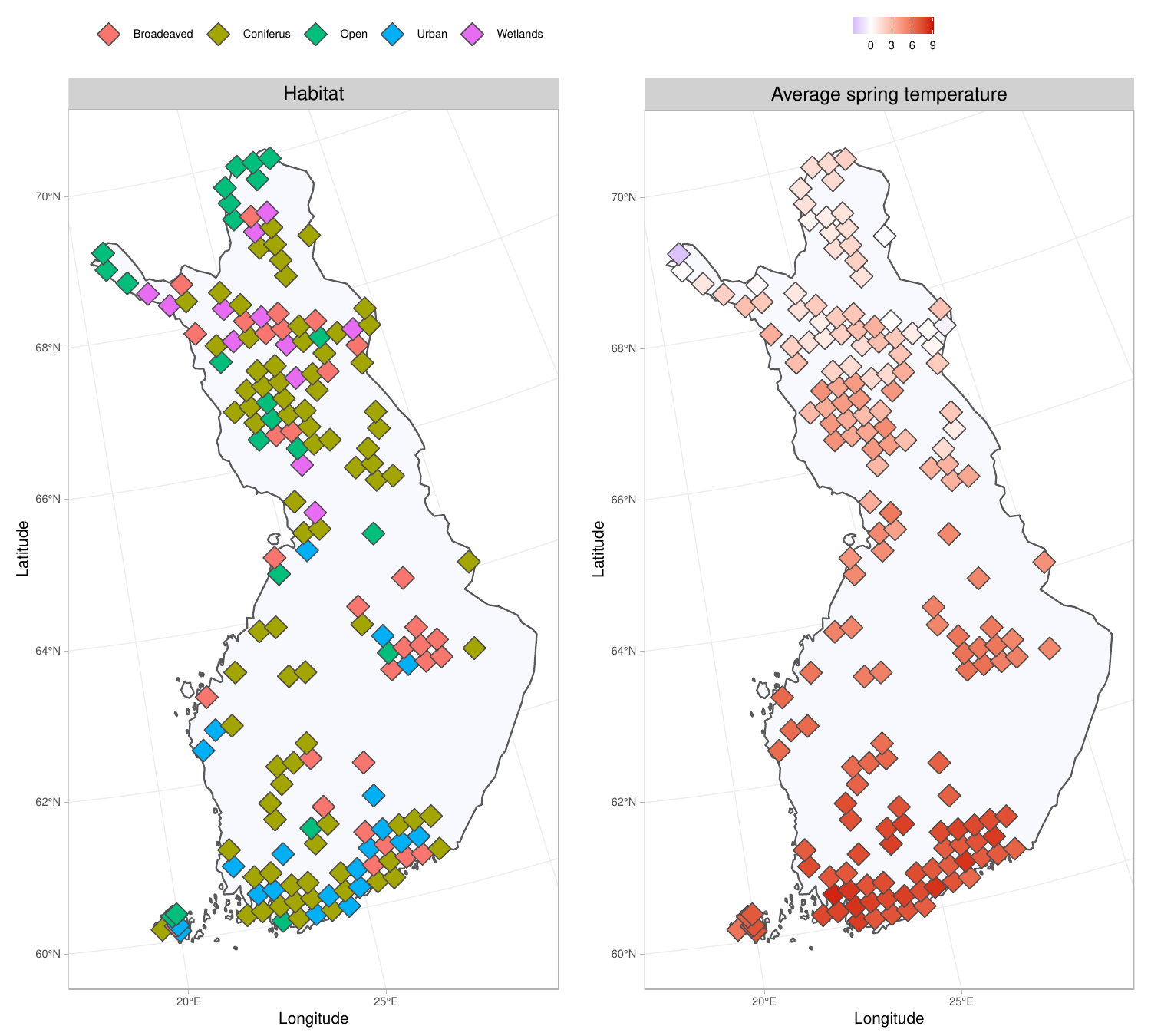}
\caption{\small 
Habitat type (left) and average temperature in April-May across 2006-2008 (right) for each site location.}
\label{fig:birds_cov}
\end{figure}

Figure \ref{fig:birds_cov} contains the summary of the environmental covariates used in Section \ref{sec:birds}

\clearpage
\bibliography{biblio}

\end{document}